\documentclass[lettersize,journal]{IEEEtran}

\usepackage{cite}
\usepackage{amsmath,amssymb,amsfonts,bm}
\usepackage{mathtools}
\usepackage{algorithmic}
\usepackage{multirow}
\usepackage{bigints}
\usepackage{multicol}
\usepackage{textcomp}
\usepackage{longtable}
\usepackage{booktabs}
\usepackage{graphicx}
\usepackage[ruled,vlined]{algorithm2e}
\usepackage{textcomp}
\usepackage{algorithmic}
\usepackage{array}
\usepackage{comment}
\usepackage{esvect}
\usepackage{subcaption}
\usepackage{xcolor}
\usepackage{dirtree}
\usepackage{caption}
\usepackage{balance}

\newcolumntype{M}[1]{>{\centering\arraybackslash}m{#1}}
\usepackage{stfloats}

\begin{document}

\title{Stochastic Geometry for Modeling and Analysis of Sensing and Communications: A Survey}

\author{Harris K.~Armeniakos,~\IEEEmembership{Member,~IEEE,} Petros S.~Bithas,~\IEEEmembership{Senior Member,~IEEE,}\\ Sotiris A. Tegos,~\IEEEmembership{Senior Member,~IEEE,} Athanasios G.~Kanatas,~\IEEEmembership{Senior Member,~IEEE,}   \\ and George K. Karagiannidis,~\IEEEmembership{Fellow,~IEEE}

\thanks{H. K. Armeniakos and A. G. Kanatas  are with the Department
of Digital Systems, University of Piraeus, 18534 Piraeus, Greece (e-mail:
\{harmen, kanatas\}@unipi.gr).}
\thanks{P. S. Bithas is with the Department of Digital Industry Technologies,
National and Kapodistrian University of Athens, 15772 Athens, Greece (email: pbithas@uoa.gr)}
\thanks{S. A. Tegos,  and G. K. Karagiannidis are
with the Aristotle University of Thessaloniki, 54124 Thessaloniki, Greece
(tegosoti@auth.gr, geokarag@auth.gr).}

}

\maketitle

\begin{abstract}
One of the most promising technologies for next-generation wireless networks is integrated communication and sensing (ISAC). It is considered a key enabler for applications that require both enhanced communication and accurate sensing capabilities. Examples of such applications include smart environments, augmented and virtual reality, or the internet of things, where the capabilities of intelligent sensing and broadband communications are vital. Therefore, ISAC has attracted the research interest of both academia and industry, and many investigations have been carried out over the past decade. The articles in the literature include system models, performance evaluation, and optimization studies of several ISAC alternative designs. 
Stochastic geometry is the study and analysis of random spatial patterns, and as such, stochastic geometry tools have been considered for the performance evaluation of wireless networks with different types of nodes. In this paper, we aim to provide a comprehensive survey of current research progress in performance evaluation of ISAC systems using stochastic geometry tools. The survey covers terrestrial, aerial, and vehicular networks, where the random spatial location of the corresponding network elements and propagation scatterers and/or blockages is treated with various point processes. The paper starts with a short tutorial on ISAC technology, stochastic geometry tools, and metrics used in performance evaluation of communication and sensing. Then, the technical components of the system models utilized in the surveyed papers are discussed. Subsequently, we present the key results of the literature in all types of networks using three levels of integration: sensing-assisted communication, communication-assisted sensing, and joint sensing and communication. Finally, future research challenges and promising directions are discussed.
\end{abstract}


\section{Introduction}
Sixth-generation (6G) networks are expected to revolutionize communications in order to realize the internet of things (IoT), enabling seamless connectivity among people, computing systems, vehicles, devices, and sensors \cite{9040264}. To support the various new use cases that are expected to emerge, e.g., augmented/virtual reality, holographic telepresence, industry 5.0, autonomous mobility, etc., 6G systems will be designed to meet stringent network requirements. For example, augmented and virtual reality applications require capacity above 1 Tb/s, while autonomous mobility requires reliability above 99.99999 percent and latency below 1 ms. In IMT-2030, it was stated that the enablers of these networks will be artificial intelligence, quantum technology, THz communications, and the integration of sensing and communication functions \cite{recommendation2023framework}. To quantify the performance of these technologies, new key performance indicators should be proposed (or  existing ones extended), related to positioning accuracy, timeliness and three-dimensional (3D) coverage \cite{jiang2021road}.

Integrated sensing and communication (ISAC) has gained increasing interest in recent years. In the past, sensing and communication functions were operated separately, requiring the use of dedicated transmitters, receivers, spectrum, and beamformers. However, in recent years, taking advantage of the enormous evolution of technology, it has become apparent that a highly efficient approach is to share these resources \cite{wymeersch2021integration}. ISAC is expected to revolutionize next-generation wireless networks by seamlessly integrating time, frequency, waveform design and hardware for both functions \cite{tan2021integrated}. This novel approach is expected to enable key applications of 6G networks such as precise localization, tracking, gesture recognition, and augmented reality. In addition, ISAC will increase data rates with improved spectral efficiency and ensure the ultra-low latency required for autonomous vehicle applications. In this context, it is desirable to jointly design the sensing and communication systems so that they can share the same frequency band and hardware to improve spectrum efficiency and reduce hardware costs. This motivates the study of ISAC. The exponential growth of ISAC-related research is also demonstrated by the numerous workshops that have been organized in flagship IEEE conferences, e.g., IEEE International Conference on Communications and IEEE Global Communications Conference, and special issues of journals \cite{special}. It should be noted that recently the IEEE has introduced an ISAC emerging technology initiative (ETI) whose purpose is to explore and support a wide variety of research directions and standardization opportunities related to ISAC \cite{ISAC_ITE}.

Stochastic geometry (SG) (also referred to as geometrical probability) is a field of applied probability that aims to provide tractable mathematical models and appropriate statistical methods, to study and analyze random phenomena in $\mathbb{R}^d$. Under the umbrella of wireless communication networks, the location of network nodes is randomly scattered over an enormous number of possibilities, where designing the system for each network realization would be time-consuming and resource-intensive \cite{baccelli1997stochastic,baccelli1998stochastic}.  The task becomes even more challenging when considering the ever-increasing complexity and heterogeneity of 6G wireless networks. Instead, SG tools can model the spatial location of network entities in the wireless environment according to point processes, thus implicitly considering all possible network realizations and generally capturing the main dependencies of the network. As a result, system-level performance analysis in terms of key performance metrics such as coverage probability and average data rate is feasible and provides meaningful insight. Although a cross-fertilization between SG and artificial intelligence can be made to provide more robust and accurate frameworks for performance evaluation \cite{8742579,8765703,8885526,saha2019machine}, one can conclude that SG is expected to remain a fundamental area of research for the foreseeable future.

\begin{table}
    \caption{List of Important Abbreviations}
    \centering
    \renewcommand{\arraystretch}{1.14}
    \begin{tabular}{|p{1.6cm}|p{6.0cm}|}
        \hline
        \textbf{Abbreviation} & \textbf{Description} \\
        \hline
        1D & One-dimensional \\
        \hline
        2D & Two-dimensional \\
        \hline
        3D & Three-dimensional \\
        \hline
        6G & Sixth-generation \\
        \hline
        AI & Artificial Intelligence \\
        \hline
        AoI & Age of Information\\
        \hline
        ASE & Area Spectral Efficiency\\
        \hline
        BPP & Binomial Point Process \\
        \hline
        BS & Base Station \\
        \hline
        CAS & Communication Assisted Sensing \\ 
        \hline
        CDF & Cumulative Distribution Function \\
        \hline
        CoMP & Coordinated Multi-point\\
        \hline
        CRLB & Cramer-Rao Lower Bound\\
        \hline
        CSMA & Carrier-Sense Multiple Access \\
        \hline
        DPP & Determinantal Cluster Process \\
        \hline
        EE & Energy Efficiency\\
        \hline
        GPP & Ginibre Point Process \\
        \hline
        HPPP & Homogeneous Poisson Point Process \\
        \hline
        IoT & Internet of Things \\
        \hline
        ISAC & Integrated Sensing and Communication\\
        \hline
        JCAS & Joint Communication and Sensing\\
        \hline
        JRDCCP & 
        Joint radar Detection and Communication Coverage
        Probability\\
        \hline
        JRSCCP & Joint radar Success and Communication Coverage Probability\\
        \hline
        JSRS &
        Joint Synchronization Signal Block and Reference Signal-based Sensing\\
        \hline
        LoS & Line-of-Sight\\
        \hline
        MCP & Mat\'ern Cluster Process \\
        \hline
        MHCP & Mat\'ern Hard-Core Point Processes\\
        \hline
        MIMO/MISO & Multiple-Input Multiple-Output/Single Output \\
        \hline
        ML & Machine Learning \\
        \hline
        mmWave & millimeter wave\\
        \hline
        MU & Mobile User\\
        \hline
        NLoS & Non-Line-of-Sight\\
        \hline
        NOMA & Non-Orthogonal Multiple Access\\
        \hline
        OFDM & Orthogonal Frequency Division Multiplexing\\
        \hline
        PDF & Probability Density Function \\
        \hline
        PGFL & Probability Generating Functional \\
        \hline
        PLP & Poisson Line Process \\
        \hline
        PMF & Probability Mass Function \\
        \hline
        PPP & Poisson Point Process \\
        \hline
        PSE & Potential Spectral Efficiency\\
        \hline
        RIS & Reconfigurable Intelligent Surface\\
        \hline
        RSU & Road-side Unit \\
        \hline
        SAC & Sensing Assisted Communication \\
        \hline
        SAR &  synthetic aperture radar\\
        \hline
        SCINR & Signal-to-Clutter-plus-Interference-plus-Noise Ratio \\
        \hline
        SCNR & Signal-to-Clutter-plus-Noise Ratio \\
        \hline
        SG & Stochastic Geometry\\
        \hline
        SIMO & Single Input Multiple Output\\
        \hline
        SINR & Signal-to-Interference-plus-Noise Ratio \\
        \hline
        SIR & Signal-to-Interference Ratio \\
        \hline
        SOMA & Spectrum Overlay Multiple Access\\
        \hline
        SSB & Synchronization Signal Block\\
        \hline
        TCP & Thomas Cluster Process \\
        \hline
        TDMA & Time Division Multiple Access \\
        \hline
        THz & Terahertz\\
        \hline
        UAV & Unmanned Aerial Vehicle\\
        \hline
        UE & User Equipment \\
        \hline
        V2X & Vehicle-to-Everything\\
        \hline
    \end{tabular}
    \label{tab:abbreviations}
\end{table}

\subsection{Relevant Surveys}
There are a number of survey papers in the literature on ISAC/JCAS and SG. In this subsection, we briefly present these papers. In \cite{10418473}, a detailed review of recent developments in ISAC systems is presented, focusing on their fundamental principles, physical layer design, networking considerations, and practical applications. In \cite{9393464}, a detailed literature review of resource management approaches for joint radar and communication systems is presented. In \cite{10812728}, the benefits, functions, and challenges of integrated sensing, communication, and computation are discussed, and the relevant technical literature is cited. In \cite{8667902}, a comprehensive review is provided for three interrelated aspects of connected automated vehicles, namely, sensing and communication technologies, human factors, and information-aware controller design. In \cite{9705498}, a survey of the fundamental limitations of ISAC is conducted to understand the gap between current state-of-the-art technologies and their ultimate performance potential. The presented analysis provides valuable information and guidance for the development of advanced ISAC technologies that can operate closer to these limits. In \cite{10756650}, a detailed review of machine learning (ML)-based ISAC systems is presented, discussing common ML and deep learning models that can be used in ISAC scenarios. In \cite{9330512}, an overview of converged 6G communication, localization, and sensing systems is presented. In \cite{9585321}, an in-depth review of systems and technologies related to joint communication and radar sensing in mobile networks is given. In \cite{10646523}, a thorough review of ISAC channel modeling approaches is presented. In \cite{9829746}, a comprehensive survey of ISAC systems is presented, highlighting the benefits of sharing spectrum, hardware, and software. In \cite{10012421}, a literature review is performed for ISAC signals in the context of mobile communication systems, covering aspects related to signal design, processing, and optimization. In \cite{9924202}, a detailed survey of ISAC techniques is performed, focusing on the waveform design perspective. In \cite{10608156}, a visionary design for an ISAC-oriented unified IoT architecture that integrates software-defined communication and super-intelligent agents is presented. Furthermore, the review focuses on ISAC technology development over the past decade, highlighting new design principles for AI-powered networks that support intelligent connectivity to strengthen ISAC security. 

In addition, a few survey and tutorial papers have focused on SG analysis of wireless communication networks. In fact, \cite{5226957} presents an overview of SG and the theory of random geometric graphs as applied to results on connectivity, capacity, outage probability, and other fundamental limits of wireless networks. In \cite{7733098}, a comprehensive tutorial on SG-based analysis for cellular networks is presented. The work presents a unified SG framework for cellular networks that serves as a tutorial. Numerical examples are also presented along with demonstrations and discussions. In \cite{6524460}, an in-depth review of the literature related to SG models for single-tier, multi-tier, and cognitive cellular wireless networks is presented. In \cite{9516701}, a comprehensive tutorial on SG analysis of large-scale wireless networks that captures the spatio-temporal interference correlation is presented. The methods for characterizing the spatio-temporal signal-to-interference ratio (SIR) correlations for different network configurations are described. In \cite{9378781}, SG models and techniques developed over the last decade for wireless network performance evaluation are presented. In the following, we outline how SG has been used to capture the properties of emerging radio access networks and to quantify the benefits of key enabling technologies. 

Although there are many survey papers that present the use of SG tools in only one functionality, e.g., either in communications or in sensing only, none of the existing surveys and tutorial papers provide an in-depth review on the statistical characterization of communication- and sensing-enabled networks by exploiting SG tools. Moreover, this is the first article to provide a comprehensive tutorial that aims to address the role of point processes and SG techniques in communication and sensing networks.

\subsection{Why Is SG Needed in ISAC Studies?}
For decades, SG has served as a powerful mathematical tool for system-level performance evaluation of complex wireless networks. By modeling the spatial locations of entities in a network through simple point processes, tractable models have been developed to gain a deeper understanding of the performance of these networks. With the advent of 6G ISAC networks, more entities have come to the forefront. In fact, due to this increase in the type of devices, the network can now include communication transmitters, radar nodes, users, and targets \cite{10574257}. With the ever-increasing complexity of 6G ISAC networks, large-scale performance evaluation of these networks to gain meaningful network-level insights has become remarkably difficult. Using SG tools, the spatial distribution of the above entities can be modeled as up to four independent point processes. In this case, the benefits are threefold: i) the effect of real-world spatial randomness of entities on ISAC network performance is considered, ii) analytical frameworks that allow system-level performance evaluation can be explored, and iii) meaningful network-level insights and design guidelines for complicated 6G ISAC networks can be drawn. However, some entities in real-world ISAC networks exhibit some spatial correlation. Fortunately, advanced point processes and SG tools can capture the spatial correlation among different point processes in complex 6G ISAC terrestrial and aerial networks. Therefore, the integration of SG tools into the ISAC paradigm has become more important than ever.    

\subsection{Contributions}
Motivated by the above observations, this review paper aims to shed light on a wide range of recent state-of-the-art works on communication and sensing enabled networks, and more specifically on: i) the exploitation of point processes in the spatial modeling of their entities and ii) the role of SG tools in their performance. While a plethora of survey works focus on JCAS and ISAC networks, no review papers focus in the SG context. In this direction, the main contributions of this paper can be summarized as follows.
\begin{itemize}
\item First, we present the fundamental knowledge of SG processes used in ISAC networks, as well as the basic concepts of sensing and communication systems. We also present the main metrics used to analytically evaluate the performance of these systems;
\item We present and compare the system models that have been studied in the literature, with the goal of highlighting the commonalities among the studies and identifying any gaps in the existing research;
\item We review and discuss the relevant technical literature in terrestrial, vehicular, and aerial communication networks for the three main types of sensing and communication networks, namely joint communication and sensing (JCAS), sensing-assisted communication (SAC), and communication-assisted sensing (CAS);
\item We discuss the key results associated with the reviewed papers and present potential research directions related to SG modeling in sensing and communications.
\end{itemize}

\subsection{Structure}
The structure of this article is illustrated in Fig. 1 and is also summarized as follows. In Section II a brief tutorial is given on the key components of SG tools, the basics of ISAC systems, and the major performance metrics used in the literature. Then, Section III provides an analysis of the technical components of the system models developed in the corresponding literature. In Sections IV and V, we review the articles dealing with terrestrial and aerial/vehicular networks, respectively. The survey follows a categorization of systems into ISACs, JCAS, and SAC.
In Section VI, we summarize the key results of the presented papers and discuss the open challenges and future research directions. Finally, in Section VII, we conclude the paper.

\begin{figure}
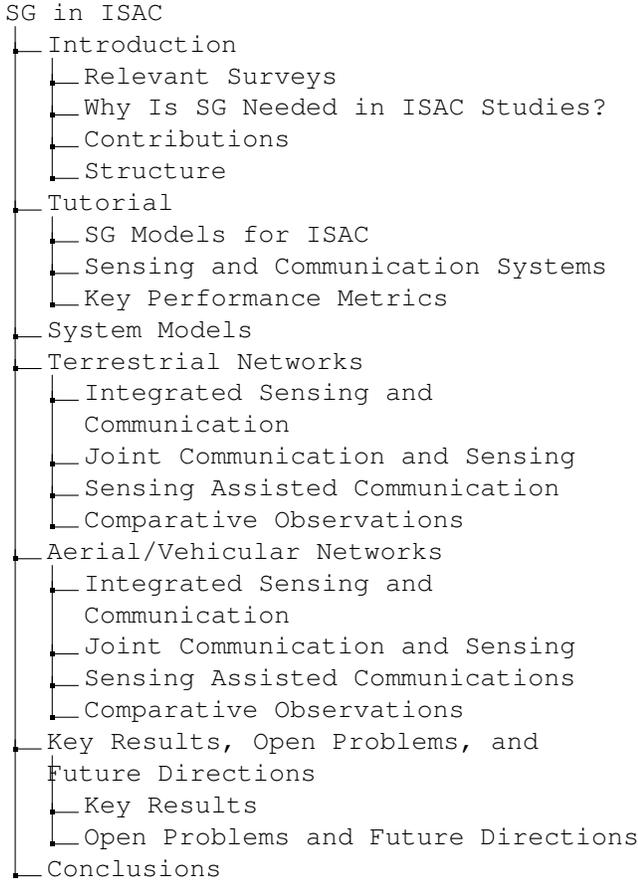

\dirtree{%
 .1 SG in ISAC.
 .2 Introduction.
 .3 Relevant Surveys.
 .3 Why Is SG Needed in ISAC Studies?.
 .3 Contributions.
 .3 Structure.
 .2 Tutorial.
 .3 SG Models for ISAC.
 .3 Sensing and Communication Systems.
 .3 Key Performance Metrics.
 .2 System Models.
 .2 Terrestrial Networks.
  .3 Integrated Sensing and Communication.
  .3 Joint Communication and Sensing.
  .3 Sensing Assisted Communication.
  .3 Comparative Observations.
 .2 Aerial/Vehicular Networks.
 .3 Integrated Sensing and Communication.
  .3 Joint Communication and Sensing.
  .3 Sensing Assisted Communications.
  .3 Comparative Observations.
  .2 Key Results, Open Problems, and Future Directions.
 .3 Key Results.
 .3 Open Problems and Future Directions.
 .2 Conclusions.
}
\caption{The structure of the paper.}
    \label{fig:organization}
\end{figure}

\section{Tutorial}
In this section, a comprehensive tutorial is presented. The aim of this tutorial is to concisely present i) fundamental concepts of communication and sensing-enabled networks and ii) mathematical preliminaries on the SG modeling in the context of communication- and sensing-enabled networks to help understand the discussions presented later in this paper. First, the tutorial defines the key spatial point processes and answers in \emph{how} and \emph{why} they have been employed in communication and sensing networks. It also introduces the key concept for statistically characterizing the aggregate interference when general two-dimensional (2D) point processes are employed. Next, the tutorial delves into the fundamental principles of ISAC networks to provide a deeper understanding of these networks. Finally, the tutorial presents communication- and sensing-based key performance metrics in the context of SG.

\subsection{Stochastic Geometry Models for ISAC}
In this subsection, key mathematical preliminaries on SG modeling are provided to help understand the discussions presented in this paper.

\subsubsection{Point Processes}
In communication- and sensing-enabled wireless networks, the spatial modeling of network nodes is abstracted to a convenient point process that captures the spatial modeling properties of the network. Therefore, based on the type of network as well as the application scenario, an appropriate point process is selected to model the spatial locations of the network entities. The most common point processes used in communication- and sensing-enabled wireless networks are defined below, along with their key properties.

\begin{figure*}[!htbp] 
\centering
\begin{subfigure}{.67\columnwidth}
\includegraphics[width=\columnwidth]{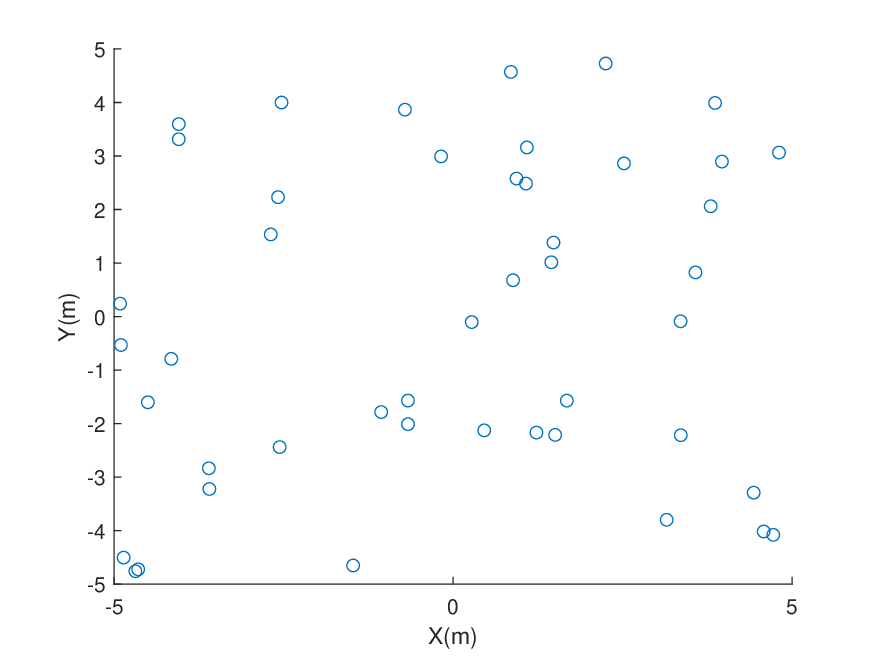}%
\caption{Realization of HPPP, $\lambda=0.5$.}%
\label{subfiga}%
\end{subfigure}
\hfill%
\begin{subfigure}{.67\columnwidth}
\includegraphics[width=\columnwidth]{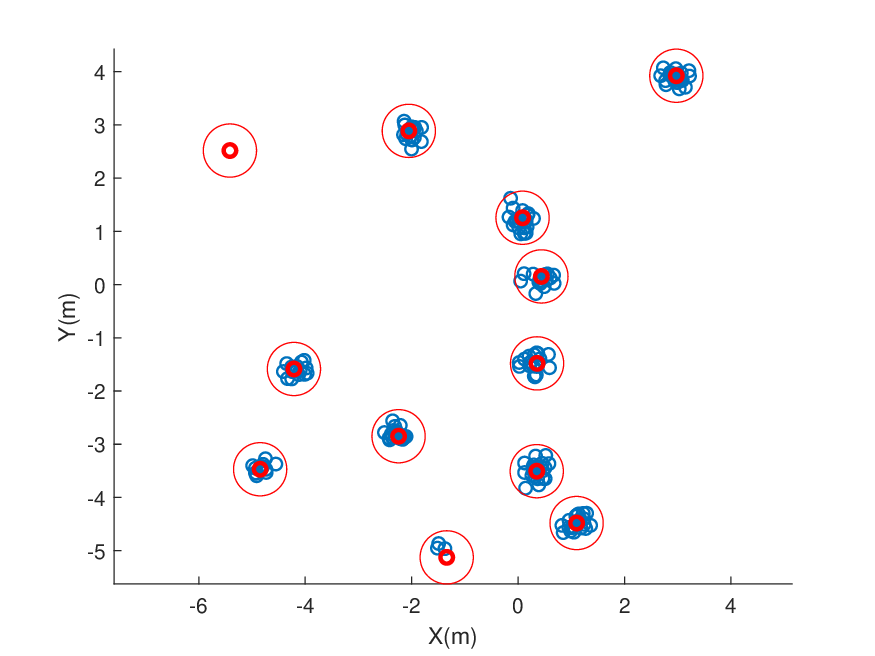}%
\caption{Realization of TCP, $\lambda_d=0.5$.}%
\label{subfigb}%
\end{subfigure}
\hfill%
\begin{subfigure}{.67\columnwidth}
\includegraphics[width=\columnwidth]{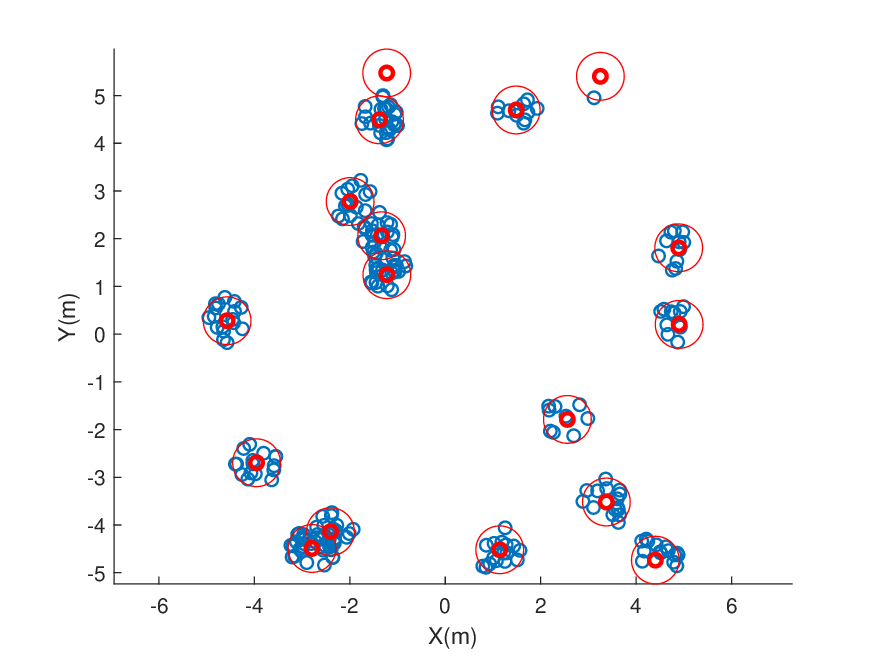}%
\caption{Realization of MCP, $\lambda_d=0.5$.}%
\label{subfigc}%
\end{subfigure}%
 \vskip\baselineskip
 \begin{subfigure}{.67\columnwidth}
\includegraphics[width=\columnwidth]{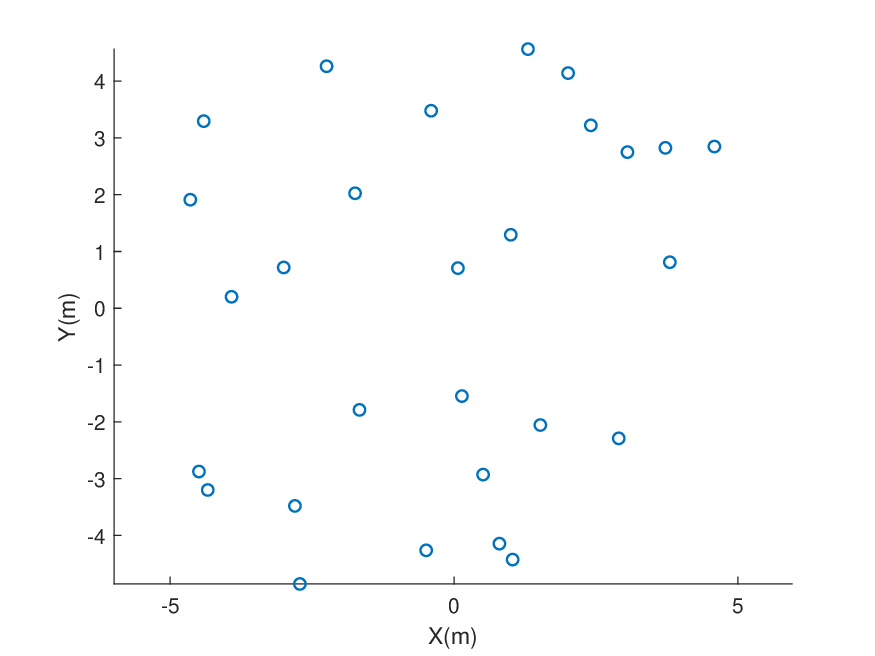}%
\caption{Realization of $\beta$-GPP, $\beta=0.5$.}%
\label{subfigd}%
\end{subfigure}%
\hfill
 \begin{subfigure}{.67\columnwidth}
\includegraphics[width=\columnwidth]{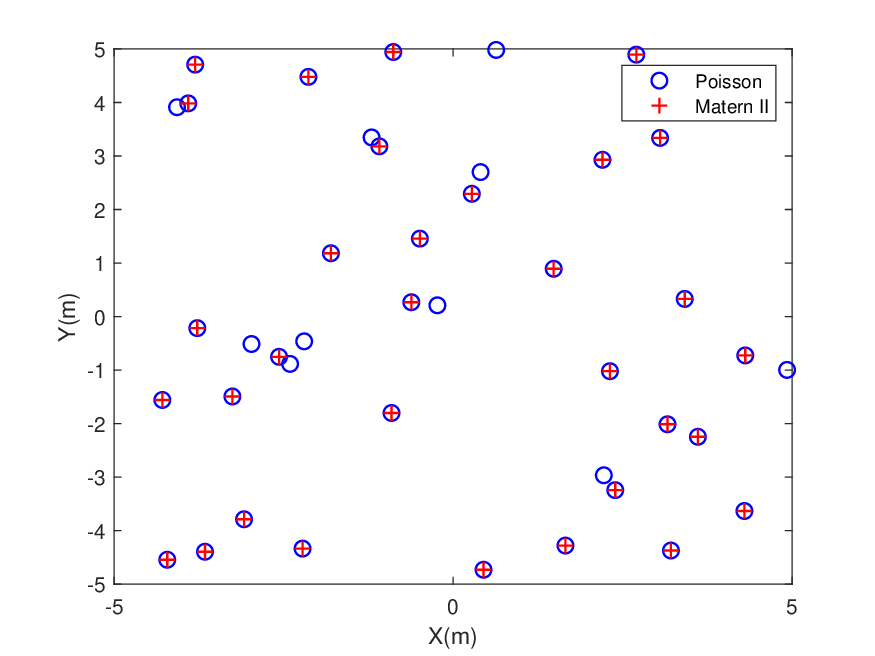}%
\caption{Realization of MHPP, $\lambda=0.5$.}%
\label{subfige}%
\end{subfigure}%
\hfill
 \begin{subfigure}{.67\columnwidth}
\includegraphics[width=\columnwidth]{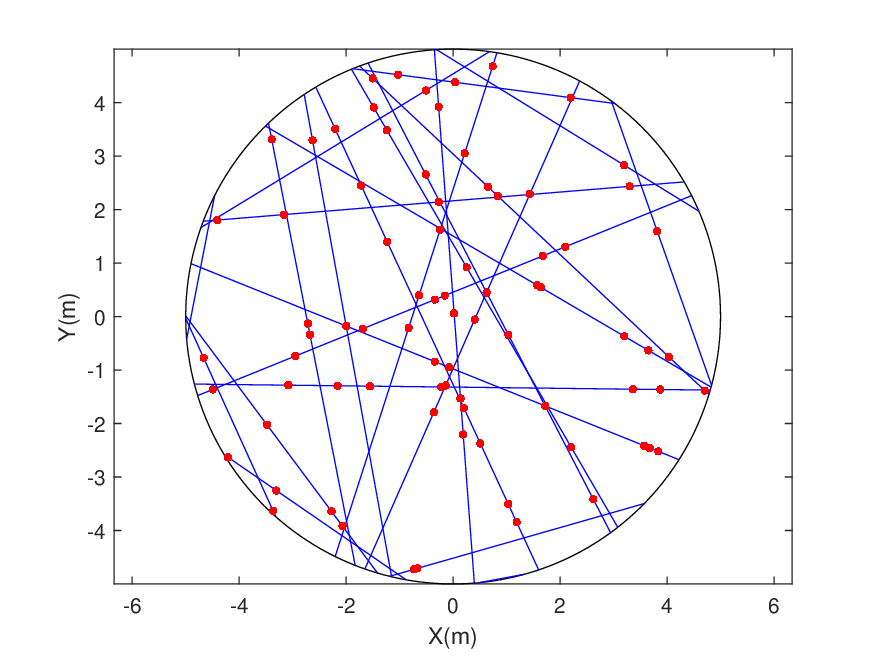}%
\caption{Realization of Cox process, $\lambda_l\hspace{-0.1cm}=\hspace{-0.1cm}\lambda_p\hspace{-0.1cm}=\hspace{-0.1cm}0.5$.}%
\label{subfigf}%
\end{subfigure}%
\caption{Realizations of point processes for communication and sensing networks.}
\label{figabc}
\end{figure*} 

\vspace{0.2cm}

\noindent \textbf{Homogeneous Poisson Point Process (HPPP):} The HPPP is considered to be the most popular point process due to its tractability and analytical flexibility and thus is widely adopted for the performance analysis of wireless networks. 

In general, an HPPP $\Psi$ of density $\lambda$ is a point process in $\mathbb{R}^d$ such that \cite{haenggi2013stochastic}: 
\begin{itemize}
  \item For every compact set $W \subset \mathbb{R}^d$, $N(W)$ has a Poisson distribution with mean $\lambda |W|$ and is characterized by a probability mass function (PMF) as 
  \begin{equation}
      \mathbb{P}[N(W)=m] = \frac{(\lambda |W|)^m}{m!} e^{-\lambda |W|}, 
  \end{equation}
  where $|\cdot|$ denotes the Lebesgue measure or the $d$-dimensional volume of the subset $W$ and $N(\cdot)$ is the counting measure, i.e., the number of points falling in $W$.
  \item If $W_1,...,W_n$ are disjointly bounded sets, then $N(W_1),...,N(W_n)$ are independent random variables. 
\end{itemize}

\textit{Motivation:} The 2D HPPP has widely been adopted in communication- and sensing-enabled wireless networks to model the spatial locations of base stations (BSs), user equipments (UEs), and sensing targets \cite{10556618,9538929,10565853} or to model the spatial locations of BSs, MTs, and blockers \cite{10633859,9893396}. In general, the PPP is used to model a network composed of a possibly infinite number of entities randomly and independently co-existing in a finite or infinite area (e.g., in a large-scale wireless network or in a cellular network). If we are interested in modeling the spatial locations of an infinite or unknown number of nodes in a finite or infinite area, the PPP is a reasonable choice. Infinite networks are usually considered for simplicity and because of the reasonable assumption that the contribution of far nodes to the aggregate interference is negligible. { Nevertheless, the HPPP is not always the best approach for realistic spatial deployment of nodes in real-world ISAC networks and one should carefully consider the trade-off between analytical tractability and accuracy when designing SG-based spatial models. }

\vspace{0.2cm}

\noindent \textbf{Matern Cluster Process (MCP):} The 2D MCP is a doubly Poisson cluster process constructed from a parent PPP $\Phi_p = \{\mathbf{x_i}\}_{i\in \mathbb{N}}$ with intensity $\lambda_p$, with each point of $\Phi_p$ substituted by a daughter cluster consisting of a PPP with an average number of points $N$ within a disk of radius $R$ centered at that point. Formally, the first-moment density for an MCP is given by $\lambda_{MCP} = N \lambda_p$. Each daughter point $\{\mathbf{y_i}\}$ is located uniformly and independently within a disk of radius $R$ around the origin. The PDF of the spatial location of each element $\{\mathbf{y_i}\}$ is given by 
\begin{equation}
f(\mathbf{y_i}) = \left\{
\begin{array}{ll}
      \frac{1}{\pi R^2}, & \|\mathbf{y_i}\|\leq R \\
     0, & \mathrm{otherwise}.\\
\end{array} 
\right.
\end{equation} 

\textit{Motivation:} When we are interested in modeling the spatial locations of nodes according to a clustering behavior around another network entity, then the MCP is a reasonable choice. To meet the requirements of URLLC and accurate detection and tracking in communication- and sensing enabled wireless networks, users and targets tend to be distributed around the transmitting nodes, making them suitable for modeling by cluster processes \cite{10412651}.

\vspace{0.2cm}

\noindent \textbf{Thomas Cluster Process (TCP):} The 2D TCP is a doubly Poisson cluster process where each cluster is a finite PPP with a Gaussian intensity function. Formally, the intensity function of a cluster is given by
\begin{equation}
\lambda_0(x) = \frac{\bar{c}}{2\pi \sigma^2}e^{-\frac{\|x\|^2}{2 \sigma^2}},
\end{equation} 
thus the  daughter points are normally distributed with variance $\sigma^2$ around each parent point, and the mean number of daughter points is $\bar{c}$. The intensity of the process is given by $\lambda_{TCP} = \bar{c} \lambda_p$.

\textit{Motivation:} The only difference between MCP and TCP is the deployment area where the nodes are randomly located. In each TCP cluster, each individual point is located according to two independent zero-mean Gaussian distributed variables with variance $\sigma^2$ describing $x$ and $y$ coordinates from the center of the cluster. In contrast, each point of an MCP is uniformly distributed on a disk. In \cite{10681925}, ISAC devices are modeled by a PPP, while the relative 2D spatial location of a sensing target clustering around each ISAC device follows an independent 2D normal distribution, thus forming a TCP.

\vspace{0.2cm}

\noindent \textbf{Determinantal Point Process (DPP):} Let $\mathbb{C}$ denote the complex plane. Then, for any function $K: \Lambda \times \Lambda \rightarrow \mathbb{C}$, we use $K(x_i,x_j)$ to denote the square matrix with $K(x_i,x_j)$ as its $(i, j)$-th entry. Formally, the point process $\Phi$ defined on a locally compact space $\Lambda$ is called a DPP with kernel $K: \Lambda \times \Lambda \rightarrow \mathbb{C}$, if its $n$-th order product density is given by
\begin{equation}
   \rho^{(n)}(x_1,...,x_n) = \mathrm{det(K(x_i,x_j))}, \quad (x_1,...,x_n) \in \Lambda^n.
\end{equation}
In this case, we denote the DPP $\Phi$  with kernel $K$ by $\Phi \sim \mathrm{DPP(K)}$. Detailed information and more general and formal definitions of DPP are presented in \cite{7155510}. In general, three DPP models have been proposed \cite{lavancier2012statistical}, namely the Gauss DPP model, the Cauchy DPP model, and the Generalized Gamma DPP model.

\textit{Motivation:} DPPs capture the repulsive behavior of nodes and produce accurate spatial models to model the locations of nodes in communication- and sensing-enabled networks when a certain correlation among the locations of nodes is required. In \cite{7155510}, DPPs were shown to outperform PPPs in predicting key performance metrics, such as coverage probability in cellular networks. In the context of communication and sensing-enabled networks, in \cite{10603243}, the Gauss DPP model was used to model the spatial locations of users in a JCAS enabled multi-user multiple input multiple output (MIMO) network. By employing the Gauss DPP, the  mitigation of communication system and radar interference was achieved while analytical tractability was maintained. In general, DPPs are an excellent choice for spatial modeling in simplistic sensing and communication system models where certain spatial correlation modeling is required. However, in practical scenarios, in which various network entities must be taken into account, e.g., terrestrial or aerial BSs, UEs, targets, and blockages, employing DPP would result to an extreme challenging analytical framework.

\vspace{0.2cm}

\noindent \textbf{$\beta$-Ginibre Point Process ($\beta$-GPP):} Let $\Phi = \{\mathbf{x_i}\}_{i\in \mathbb{N}}$, be a $\beta$-GPP with density $\lambda$ and repulsion parameter $\beta$, i.e., $\Phi \sim \mathrm{GPP(\lambda,\beta)}$. Then, $\Phi$ is a DPP with the kernel given by
\begin{equation}
    \mathcal{K}_{\beta,\lambda}(x,y) = \lambda e^{-\frac{\pi \lambda |x-y|^2}{2 \beta}},
\end{equation}
where $x,y \in \mathbb{C}$. Note that $\mathcal{K}_{\beta,\lambda}(x,y)$ represents the interaction force among the points of the process. 

The $\beta$-GPP represents a repulsive point process in which the parameter $\beta \in (0, 1]$ can be used to smoothly approach the PPP from the GPP. More specifically, when $\beta \rightarrow 1$, the points in the $\beta$-GPP experience strong repulsive behavior. On the other hand, when $\beta \rightarrow 0$ or $\lambda \rightarrow \infty$, the correlation among the points disappears, which corresponds to the case where the locations of the points are independent, i.e., the PPP case. 

\textit{Motivation:} When we are interested in modeling the spatial locations of entities with a certain repulsive behavior, then $\beta$-GPP is a reasonable choice. In fact, in real-world network deployment scenarios, BS locations are typically spatially correlated and exhibit repulsive behavior \cite{9420372}. In this case, $\beta$-GPP is a fairly tractable model for random points with spatial repulsion.
 
\vspace{0.2cm} 

\noindent \textbf{Cox Point Process:} Let $Z = \{Z(x): x \in \mathbb{S}\}$ be a non-negative random field such that with probability one $x \rightarrow Z(x)$ is a locally integrable function. Then we say that a point process $\Phi$ defined on some underlying space $\mathbb{S}$ is a Cox point process driven by $Z$ if the conditional distribution of $\Phi$ is a PPP with intensity function $Z$. 

In practice, a Cox point process for modeling the spatial locations of entities in a vehicular network is constructed as follows. We first model the spatial distribution of road systems using a motion-invariant PLP $\Phi_L$ with line intensity $\mu_L$. We then model the spatial locations of wireless nodes, including vehicular nodes and RSUs, on each line (road) using a one-dimensional (1D) HPPP with density $\lambda_n$. Now, we can say that the set of spatial locations of vehicular nodes and RSUs is modeled by the Cox processes $\Phi_r$ and $\Phi_t$, respectively, and driven by the same PLP $\Phi_L$.

\textit{Motivation:} The Cox point process is a particularly well-known point process for modeling the spatial locations of vehicles and/or RSUs in vehicular networks. In the context of vehicular networks enabled by communication and sensing, the Cox point process has already been used in \cite{9814629}. Although the Cox point process is the most reasonable model for spatially modeling entities in a vehicular network because it builds on the Poisson line process for modeling the roads and highways of vehicular networks, it usually results in intractable expressions for key performance metrics. {Therefore, the feasibility of a large-scale performance evaluation of ISAC vehicular networks at system-level is dubious.}  

\vspace{0.2cm}

\noindent \textbf{Mat\'ern Hard-Core Point Processes (MHCP) of type II:}  Starting with a basic uniform PPP $\Phi_b$ with intensity $\lambda_b$, add to each point $x$ an independent random variable $m(x)$, called a mark, uniformly distributed in $[0,1]$. Flag to remove of all points that have a neighbor within distance $r$ that has a smaller mark. Then remove all flagged points. Formally,
\begin{equation}\label{MHCP}
\Phi_h \triangleq  \{x \in \Phi_b:m(x)<m(y)\,\,  \mathrm{for\,\, all}\,\, y \in \Phi_b \cap b(x,r) \setminus \{x\}\},
\end{equation}
where $b(x,r)$ denotes a 2D sphere of radius $r$ centered at $x$. Assume that $\Phi_h$ is a 2D MHCP. To determine the intensity $\lambda_h$ of $\Phi_h $, we first condition on a point with a mark $t$. This point is retained with probability $e^{-t \lambda_b  |b(0,r)|}$, since $t \lambda_b$ is the density of points with marks smaller than $t$. Deconditioning over $t$ yields
\begin{equation}
\lambda_h = \lambda_b \int_0^1 e^{-t \lambda_b  |b(0,r)|} \mathrm{d}t = \frac{1-e^{-t \lambda_b  |b(0,r)|}}{|b(0,r)|},
\end{equation}
where $|.|$ denotes the Lebesgue measure.  

\textit{Motivation:} Hard-core processes are point processes for modeling the spatial locations of nodes, where the nodes are forbidden to be closer than a certain minimum distance. The key idea for building hard-core point processes and achieving the minimum distance constraint is to start with a conventional PPP with no restriction and then remove the points that violate the minimum distance condition. In the context of communication- and sensing-enabled networks, the type II 1D MHCP has been used in \cite{10582835} to model the spatial locations of each vehicle, where a minimum safety distance between vehicles is ensured. The 1D MHCP serves as an alternative and analytically tractable choice to model vehicles in ISAC-enabled vehicular networks compared to the Cox point process. In fact, in \cite{10582835}, the signal-to-interference-plus-noise ratio (SINR)-based coverage probability is derived in terms of a single first-order Marcum-Q function.

\subsubsection{Statistical Characterization of Aggregate Interference for General 2D Point Processes}

With the ever-increasing complexity of communication- and sensing-enabled wireless networks and the vast number of novel smart applications that these networks now support, the statistical characterization of the aggregate interference power distribution becomes extremely difficult. This is because the aggregate interference now includes many random factors such as radar cross section (RCS), antenna gains, etc. Due to the many random terms, it is not feasible to statistically characterize the aggregate interference by directly building on the distribution for the sum and product of random variables. Instead, the most widely adopted techniques for characterizing the aggregate interference power distribution are Campbell’s theorem and the probability generating functional (PGFL), which are defined for the practical case of 2D point processes.  

\begin{itemize}
    \item \textbf{Campbell Theorem:} Let $\Phi$ denote a general point process in $\mathbb{R}^2$ and $f:\mathbb{R}^2\rightarrow\mathbb{R}$ be a measurable function. Then, 
    \begin{equation}\label{Campbell}
        \mathbb{E}\Bigg\{ \sum_{x_i \in \Phi} f(x_i) \Bigg\} = \int_{\mathbb{R}^2} f(x) \lambda(x) \mathrm{d} x,
    \end{equation}
where $\lambda(x)$ denotes the intensity function of the 2D point process $\Phi$. 

 \item \textbf{PGFL:} For $u(x) \in [0,1]$ and $\int_{\mathbb{R}^2} (1-u(x))\mathrm{d} x < \infty$, the PGFL for the point process $\Phi$ is given by
     \begin{equation}\label{PGFL}
        \mathbb{E}\Bigg\{ \prod_{x \in \Phi} u(x) \Bigg\} = \exp\Big( -\int_{\mathbb{R}^2} (1-u(x)) \lambda(x) \mathrm{d} x\Big).
    \end{equation}
\end{itemize}

For the conventional case of a PPP, the derivation of the characteristic function of the aggregate interference using the PGFL usually yields analytically tractable results. This is because according to Slivnyak’s theorem, the reduced Palm distribution of the PPP is the same as its ordinary distribution. In other words, by removing a point from the PPP, the remaining process is the same as the original point process with the same intensity. Therefore, the PGFL for the PPP can also be expressed as \eqref{PGFL}.

\subsection{Sensing and Communication Systems}
\begin{figure*}
  \centering
    \includegraphics[width=7in]{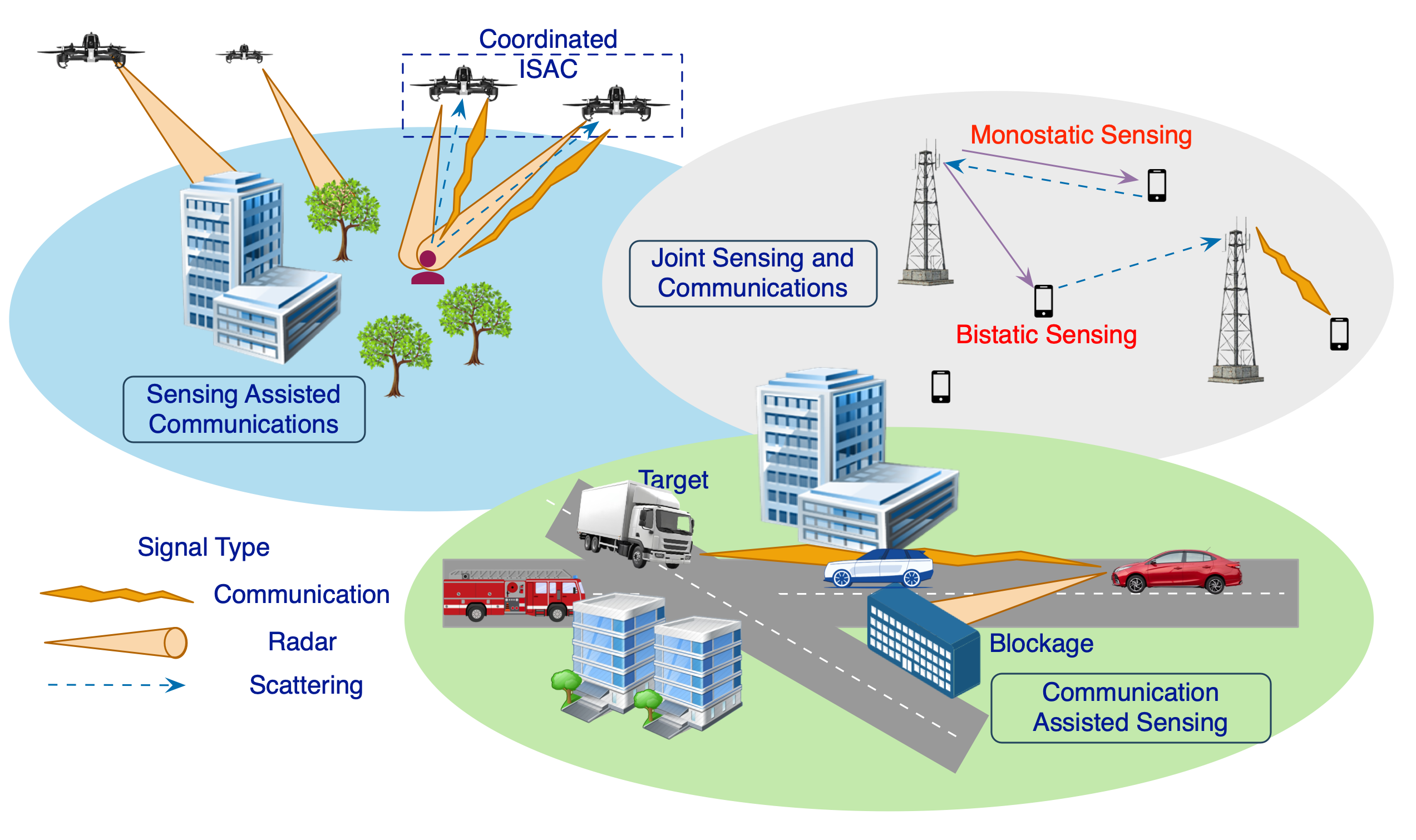}
    \caption{Types and configurations of sensing and communication networks.}
    \label{fig:system}
\end{figure*}

The driving forces for investigating the coexistence of sensing and communication are the result of two independent processes \cite{8828016}: 
\begin{enumerate}
    \item The technological advances that have made it possible to use very high carrier frequencies for communication purposes, traditionally used in radar applications, in conjunction with the massive antenna arrays that are currently available;
    \item The demanding requirements for reduced hardware costs and power consumption imposed by the upcoming 6G networks. 
\end{enumerate}
As a result of the convergence of these technologies, several benefits are expected, the most important of which are the reduction of electromagnetic pollution, efficient use of power and spectrum, improved wireless communication and sensing performance, and the introduction of novel applications. According to the research priorities and architectures studied, three main types of ISAC networks have been identified \cite{wu2022joint}, which are also illustrated in Fig. \ref{fig:system}.

\begin{itemize}
    \item Sensing-assisted communication (SAC);
    \item Communication-assisted sensing (CAS);
    \item Joint communication and sensing (JCAS).
\end{itemize}
The main objective of SAC is to increase the reliability of communication links by using environmental information obtained from radar sensing. This approach aims at significantly improving of the channel estimation accuracy while reducing the signaling overhead and identifying blockages. The additional information provided by sensing can be used in various communication techniques such as beam training, beam alignment, power adaptation, especially in high-mobility environments \cite{10845869}. 

CAS focuses on how the sensing functionality can be enhanced by exploiting the operation of a communication system. For example, communication signals reflected from objects can be used to detect or track objects without the need for additional sensors. This approach provides another degree of freedom in scenarios where sensing must be performed under non-line-of-sight (NLoS) propagation conditions. In JCAS, a single transmitted signal is used for both communication and sensing functionalities, an approach that is highly efficient and cost-effective. The implementation of JCAS technologies and services in a multi-functional network would be one of the pillars in a 6G network that is expected to transform wireless communication by providing ultra-high data rates, extremely low latency, massive device connectivity, and ultra-high reliability and availability. Clearly, in JCAS networks, the main objective is to design and optimize the signal waveform, system and network architecture equally for both functions. This approach is in contrast to others, i.e., CAS, SAC, in which the research focus is on how the secondary functionality can be used to improve the primary functionality, without significantly affecting its operating principles. 

The level of integration of sensing and communication systems can vary from simple coexistence to deeper integration. Specifically, at the lowest levels, communication and sensing share locations, but not spectrum or hardware. Deeper integration levels involve sharing spectrum and hardware, taking advantage of the existing communication infrastructure for sensing. Further integration involves scenarios where the same waveform is used for both functions. Moreover, in even deeper integration, communication signals are reused for sensing, reducing overhead and energy consumption. 

Regarding the physical placement of the transmitter and receiver for sensing, various types of sensing configuration have been proposed: monostatic, bistatic, monostatic sensing with coordination, bistatic sensing with coordination \cite{10422712}, which are also illustrated in Fig.~\ref{fig:system}. In monostatic phased-array radar, the transmit and receive arrays are co-located, often utilizing the same antenna array for both transmission and reception. A key challenge in this scenario is related to interference from the transmit array affecting the receive array, which must be mitigated \cite{9705498}. In bistatic phased-array radar, the transmit and receive arrays are located at separate locations, and thus reduced interference is expected. However, additional information is required because the angles of departure and arrival differ. In coordinated monostatic sensing, a sensing organizer, such as an access point, manages and synchronizes the transmissions of one or more devices performing monostatic sensing. The key challenges in this case include mutual interference management, synchronization, and scalability issues. Similarly, in coordinated bistatic sensing, the sensing organizer coordinates multiple sensing receivers and ensures that their operations are effectively coordinated \cite{10422712}. In this case, in addition to the challenges of coordinated monostatic sensing, increased computational overhead must be supported.

\subsection{Key Performance Metrics}
In this subsection, the most important key performance metrics are presented that have been used in the context of sensing and communication analysis using SG tools.
\begin{itemize}

\item \textbf{Communication Coverage Probability:} This is the most common performance metric used in communication- and sensing-enabled networks. It is also referred to as \emph{success probability}. It is formally defined as the probability that the communication SINR at the receiving node exceeds a target threshold $\gamma$ required for successful demodulation of the communication signal. Mathematically, it is expressed as
\begin{equation}\label{pcov}
    P_c =  \mathbb{P}[\mathrm{SINR_{com}}>\gamma]. 
\end{equation}
The SINR threshold $\gamma$ is set with respect to the achievable communication capacity requirement, given by $B \log_2(1+\mathrm{SINR_{com}})$, where $B$ is the transmission bandwidth. The definition in \eqref{pcov} has been used in \cite{10562219,10569084,10633859,9420372,10568512,10433485,10320397}.

\item \textbf{Communication Average Data Rate:} Let $\Phi_\mathrm{bs}$ denote the PPP that models the spatial locations of BSs with density $\lambda_{bs}$. Then, the communication average data rate is given by $R_c = \mathbb{E}_{\Phi_\mathrm{bs}}[\log(1+{\mathrm{SINR_{com}}})]$, where $\mathbb{E}_{\Phi_\mathrm{bs}}[\cdot]$ denotes expectation with respect to the point process $\Phi_\mathrm{bs}$. 

\item \textbf{Communication Area Spectral Efficiency (ASE):} Let $K$ denote the number of users served in each cell. Then, the communication ASE is mathematically expressed as $T_c^{\mathrm{ASE}} = \lambda_{bs} K R_c$ \cite{10735119}.

\item \textbf{Sensing Average Data Rate:} Let $\Phi_\mathrm{bs}$ denote the PPP that models the spatial locations of the BSs. Then, the target’s average radar information rate based on the sensing SINR ${\mathrm{SINR_{sen}}}$ is given by $R_s = \mathbb{E}_{\Phi_\mathrm{bs}}[\log(1+{\mathrm{SINR_{sen}}})]$.  

\item \textbf{Sensing ASE:} Let $J$ denote the number of targets detected in each cell. Then, the sensing ASE is mathematically expressed as $T_c^{\mathrm{ASE}} = \lambda_{bs} J R_s$ \cite{10735119}.

\item \textbf{Potential Spectral Efficiency (PSE):} A key performance metric for the design of ISAC networks is the PSE, which is the network information rate per unit area (measured in bps/m$^2$) that corresponds to the minimum tolerated signal quality for reliable transmission.

\begin{enumerate}
    \item  \emph{Communication PSE:} Conventionally, the communication PSE of the communication system is defined as 
    \begin{equation}\label{commPSE}
    R_c =  \lambda_{bs}\log_2(1+\gamma_c)\mathbb{P}[\gamma_0>\gamma_c],
\end{equation}
where $\gamma_c$ denotes a target threshold for reliable communication and $\gamma_0$ is the received SINR at the typical UE. 

 \item  \emph{Radar PSE:} The communication PSE of the communication system is defined as 
    \begin{equation}\label{radarPSE}
    R_r =  \lambda_{bs}\log_2(1+\gamma_r)\mathbb{P}[\gamma_i>\gamma_r],
\end{equation}
where $\gamma_r$ is a target threshold for reliable decoding and $\gamma_i$ is the received SINR at the $i$-th BS.
\end{enumerate}
In the context of ISAC networks, the definitions in \eqref{commPSE} and \eqref{radarPSE} have been used in \cite{10490156}.

\item \textbf{Energy Efficiency (EE):} EE is a key performance metric in communication and sensing enabled networks and typically accounts for both the spatial locations of network entities and the number of these entities. Therefore, it is of great research interest to optimize the density and spatial locations of nodes to maximize the energy efficiency of the network. The EE is defined as a function of the communication and sensing rates and is given by 
\begin{equation}\label{EE}
    EE = \frac{R_c+R_r}{P},
\end{equation}
where $P$ denotes the total consumed power. In the context of ISAC networks, the definition in \eqref{EE} and \eqref{radarPSE} has been employed in \cite{10490156}.

\item \textbf{JCAS Coverage Probability:} Let $\Phi_\mathrm{UE}$ and $\Phi_\mathrm{s}$  denote the PPPs modeling the spatial locations of UEs and sensing objects, respectively.  Formally, the JCAS coverage probability is defined as the \textit{joint} fraction of UEs and sensing objects whose corresponding SINR is above some corresponding threshold. Mathematically, 
\begin{equation}\label{jcascov}
\begin{split}
    &P_{c,\mathrm{JCAS}}(\gamma_{\mathrm{com}},\gamma_{\mathrm{sen}})= \\
    &\frac{\lambda_\mathrm{UE} \mathbb{P}^{0}_{\Phi_\mathrm{UE}}[\mathrm{SINR_{com}}\geq \gamma_{\mathrm{com}}]}{\lambda_\mathrm{UE}+\lambda_\mathrm{s}} \!+\! \frac{\lambda_\mathrm{s} \mathbb{P}^{0}_{\Phi_\mathrm{s}}[\mathrm{SINR_{sen}}\geq \gamma_{\mathrm{sen}}]}{\lambda_\mathrm{UE}+\lambda_\mathrm{s}},
\end{split}
\end{equation}
where $\mathbb{P}^{0}_{\Phi_\mathrm{s}}$ and $\mathbb{P}^{0}_{\Phi_\mathrm{UE}}$  denote the Palm measures associated with the point processes $\Phi_\mathrm{s}$ and $\Phi_\mathrm{UE}$, respectively. In the context of JCAS enabled networks, this definition has been employed in \cite{10320397} and in \cite{10615428} to characterize the meta-distribution. 

\item \textbf{JCAS Ergodic Rate:} The JCAS ergodic rate is defined as the \textit{joint} spatial average of the corresponding rate functions. Mathematically, JCAS ergodic rate is expressed as 
\begin{equation}\label{jcasrate}
\begin{split}
    E_{c,\mathrm{JCAS}} & = \frac{\lambda_\mathrm{UE} }{\lambda_\mathrm{UE}+\lambda_\mathrm{s}}\mathbb{E}^{0}_{\Phi_\mathrm{UE}}[\log(1+\mathrm{SINR_{com}})]\\
    & \quad + \frac{\lambda_\mathrm{s} }{\lambda_\mathrm{UE}+\lambda_\mathrm{s}}\mathbb{E}^{0}_{\Phi_\mathrm{s}}[\log(1+\mathrm{SINR_{sen}})],
\end{split}
\end{equation}
where $\mathbb{E}^{0}_{\Phi_\mathrm{UE}}[\cdot]$ and $\mathbb{E}^{0}_{\Phi_\mathrm{s}}[\cdot]$ denote expectation with respect to the point process $\Phi_\mathrm{UE}$ and $\Phi_\mathrm{s}$, respectively. This definition has been used in \cite{10320397}.

\item \textbf{$\mathcal{L}$-conditional Positioning Coverage Probability:} Under the condition of $\mathcal{L}$
BSs participating in the positioning process of an ISAC-enabled network, the positioning coverage is defined as the probability that the received signal strength-based CRLB $\underline{\mathcal{C}}$ is not greater than
the threshold value $\epsilon$. Mathematically, the $\mathcal{L}$-conditional positioning coverage probability is defined as 
\begin{equation}
    P_p(\epsilon|\mathcal{L}) =  \mathbb{P}[\underline{\mathcal{C}}\leq\epsilon].
\end{equation}

\item \textbf{$\mathcal{L}$-localizability Probability:} The $\mathcal{L}$-localizability in an ISAC network can be defined as the probability that the SINR of the $\mathcal{L}$-nearest BS is greater than the threshold $\gamma$. In other words, the $\mathcal{L}$-localizability probability is defined  as the probability that at least $\mathcal{L}$ BSs participate in the positioning process. Based on a simple SINR metric, it is mathematically expressed as 
\begin{equation}
\begin{split}
    &P_L(\mathcal{L}|\gamma)= \mathbb{P}\Bigg[\frac{P_t h_\mathcal{L} r_\mathcal{L}^{-\beta}}{\sum_{i=\mathcal{L}+1}^{\infty}P_t h_i r_i^{-\beta}+N_0}\geq\gamma\Bigg],
\end{split}
\end{equation}
where $h_{(\cdot)}$ denotes the small-scale fading channel coefficient power gain, $\beta$ denotes the path-loss exponent $P_t$ is the transit power of the BSs and $N_0$ denotes the average power of received noise. 

\item \textbf{Positioning Coverage Rate:} The coverage rate in the positioning process of an ISAC-enabled network is defined as
\begin{equation}
\begin{split}
    P_p(\epsilon) & = P_p(\epsilon|L_P)P_L(L_P|\gamma) + \sum_{l=3}^{L_P-1}P_p(\epsilon|l)f_L(l|\gamma)\\
    & \quad +(1-P_L(3|\gamma))u(\epsilon-N_L),
\end{split}
\end{equation}
where $L_P$ denotes the maximum number of BSs that can participate in the positioning process, $u(\cdot)$ denotes the  step function, $N_L$ denotes a sufficiently large value and $f_L(\mathcal{L})$ is the PMF of exact $\mathcal{L}$ BSs involved in the positioning process. It should be also noted that the case where $l<3$ corresponds to the unlocalizable scenario, as shown in \cite{10556618}. 

\item \textbf{Communication Coverage Rate:} The coverage rate in the communication process of an ISAC-enabled network is defined as  
\begin{equation}\label{communcoverage}
    P_c(\epsilon_c) = \mathbb{P}[\mathrm{SINR}_{\mathrm{com}}\geq\epsilon_c], 
\end{equation}
where $\epsilon_c$ denotes a communication coverage rate target threshold. 

\item \textbf{ISAC Coverage Rate:} The coverage rate of the ISAC process in an ISAC-enabled network is defined as
\begin{equation}\label{isacoverage}
    P_{p\&c}(\epsilon,\epsilon_c) = \mathbb{P}[\underline{\mathcal{C}}\leq\epsilon,\mathrm{SINR}_{\mathrm{com}}\geq\epsilon_c].  
\end{equation}
Considering the coupling effects between the positioning and communication process of an ISAC-enabled network, $P_{p\&c}(\epsilon,\epsilon_c)$ can capture the probability that the positioning and communication bounds are jointly lower and higher than $\epsilon$ and $\epsilon_c$ at any point in the ISAC network, respectively. In the context of ISAC enabled networks, localization-based performance metrics have been used in \cite{10556618}. 

\item \textbf{Rate Coverage Probability:} The rate coverage probability is an SINR-based metric and is defined as the probability that the downlink rate of the typical UE is above a target rate threshold $\tau_r$. Mathematically, it is expressed as 
\begin{equation}\label{ratecovprob}
    P_{rate}(\tau_r) = \mathbb{P}[\frac{K}{K_0}B \log_2(1+\mathrm{SINR}) \geq \tau_r],
\end{equation}
where $K_0$ is the number of UEs in the coverage area of the typical BS and $K$ denotes the  number of
associated UEs of the typical BS. Note that the definition is conditioned on the existence of $K_0$ UEs in a given coverage area. 
The definition in \eqref{ratecovprob} has been employed in \cite{10747175}.

\item \textbf{Sensing Outage Probability:} The sensing outage probability is defined as the probability that a target cannot be successfully detected by the BS. In other words, signal-to-clutter-plus-noise ratio (SCNR) cannot exceed a certain threshold required for successful target detection. The definition has been used in \cite{10747175}.

\item \textbf{Radar Detection Coverage Probability:} The radar detection coverage probability is a key performance metric employed in most JCAS- and ISAC-related scenarios. The performance metric can be formulated based on i) SCNR, ii) signal-to-clutter-plus-interference-plus-noise ratio (SCINR), and iii) SINR. 

radar detection coverage probability is defined as the probability that the SCNR  from a radar target is above a predefined threshold $\gamma$. The metric is used to capture the likelihood of a radar target being detected by a monostatic or bistatic radar based on the SCNR metric. Mathematically, 
\begin{equation}\label{radardetection1}
    P_{\mathrm{rm}}(\gamma) = \mathbb{P}[\mathrm{SCNR}\geq\gamma].  
\end{equation}
In the context of JCAS networks, the definition in \eqref{radardetection1} has been used in \cite{ram2022optimization}. A definition similar to the one in \eqref{radardetection1} which is referred to as successful detection probability, has also been used in \cite{10747175}. In \cite{10562219}, the definition shown in \eqref{radardetection1} is referred to as sensing coverage probability. 

The definition of the radar detection coverage probability can be extended by further considering the aggregate interference from nodes with the JCAS network, i.e., an SCINR-based metric. In this case, the radar detection coverage probability can be defined as the probability that SCINR is above a predefined threshold $\gamma$, i.e.,
\begin{equation}\label{radardetection2}
    P_{\mathrm{rm}}(\gamma) = \mathbb{P}[\mathrm{SCINR}\geq\gamma].  
\end{equation}
The definition presented in \eqref{radardetection2} has been used in \cite{10433485}.  

However, considering the clutter interference in performance analysis usually makes the statistical characterization of aggregate interference intractable, even if the PPP is used to model the spatial locations of entities in a JCAS network. When considering an appropriate point process for modeling a cluttered environment, such as a \emph{cluster point process}, the calculation of the Laplace transform of the aggregate interference power distribution may result in high analytical complexity. To overcome this issue and maintain analytical tractability, the clutter interference can be considered negligible compared to the aggregate interference from nodes. In this case, the radar detection probability is defined as the probability that the radar receiver detects a real target. Equivalently, this means that the SINR-based metric for the radar receiver should exceed a predefined threshold necessary to detect the target, given that the target actually exists. 

In the context of JCAS networks, the SINR-based radar detection probability has been used in \cite{10569084,10570658,9420372,9931437}. In \cite{10217344}, the SINR-based radar detection probability was alternatively referred to as successful ranging probability. In \cite{10412651}, the above definition is referred to as successful sensing probability.    

\item \textbf{Radar False Alarm Probability:} The radar false alarm probability is defined as the probability that the radar receiver falsely detects a target that does not exist. In the context of JCAS networks, a definition similar to the one above has been used in \cite{10569084,9893396}.
  \begin{figure*}[t!]
  \centering
    \includegraphics[width=7in]{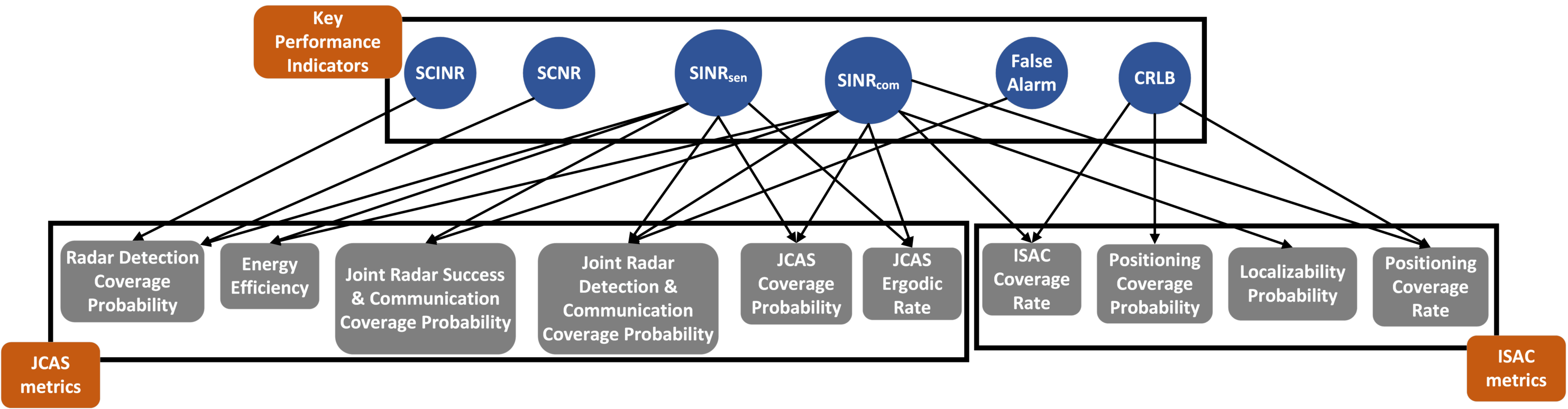}
    \caption{Major key performance indicators and metrics.}
    \label{fig:metrics}
\end{figure*}

\item \textbf{Joint radar Detection and Communication Coverage Probability (JRDCCP):} JRDCCP has been proposed as an extension of the definition of the communication coverage and the radar detection probabilities for JCAS networks. Formally, JRDCCP is defined as the probability of detecting a target with a given false alarm probability at the radar function while achieving a sufficient communication SINR.  The key idea here is that interference from communication vehicles affects also the radar function and vive versa. Assume a communication SINR threshold $\theta$ and a threshold $\gamma$ related to the radar function. $\gamma$ related to the radar function. Then, the mathematical definition of JRDCCP is given by
\begin{equation}\label{JRDCCP}
    P_{\mathrm{JRDCCP}}(\theta,\gamma) = \mathbb{P}[\mathrm{\mathrm{Pr}_s+\mathrm{Pr}_c+I}\geq\theta, \frac{\mathrm{Pr}_c}{\mathrm{Pr}_s+I}\geq\gamma],
\end{equation}
for a given false alarm probability $\mathcal{P}_{FA}(\theta) = \mathbb{P}[\mathrm{Pr}_c + I \geq \theta]$, where $\mathrm{Pr}_c$ denotes the power for the communication link, $\mathrm{Pr}_s$ denotes the power for the radar link and $I$ denotes the interference power from vehicles in the opposite direction of the road. The definition considers the probability of radar detection for a false alarm probability, while ensuring a sufficient SINR for the communication function. The threshold $\gamma$ is selected to meet the false alarm probability requirement, and the JRDCCP metric is then obtained for a given communication SINR threshold $\theta$. The definition  in \eqref{JRDCCP} has been used in \cite{10569084}.

\item \textbf{Joint radar Success and Communication Coverage Probability (JRSCCP):} JRSCCP has been proposed as an extension to the definition of success probability for JCAS networks. Formally, JRSCCP is defined as the probability of achieving simultaneously sufficient SINRs on both radar and communication functions.
Mathematically, it is expressed as
\begin{equation}\label{JRSCCP}
    P_{\mathrm{JRSCCP}}(\theta,\theta') = \mathbb{P}[\frac{\mathrm{Pr}_c}{\mathrm{Pr}_s+I}\geq\theta, \frac{\mathrm{Pr}_s}{\mathrm{Pr}_c+I}\geq\theta' ],
\end{equation}
where $\theta$ and $\theta'$ denote the communication and radar SINR target thresholds, respectively. The definition  in  \eqref{JRSCCP} has been used in \cite{10569084}. In \cite{10695883}, the definition in \eqref{JRSCCP} is employed but is referred to as ISAC coverage rate. However, the CRLB metric of the ISAC network is not considered and therefore the definition in  \eqref{JRSCCP} differs significantly from the definition presented in \eqref{isacoverage}.

\item \textbf{Beam Misalignment Probability:} According to 5G NR, the key task of beam alignment is to establish an appropriate beam pair before data transmission \cite{dahlman20205g}. The candidate beam pair is determined at the BS side based on the synchronization signal block (SSB) measurement reported by the user. Missed reception of SSBs and untimely beam switching will cause beam misalignment. Therefore, beam misalignment refers to the cases where the beams are not aligned at the UE and BS sides. In the context of ISAC networks, beam misalignment can be caused by:
\begin{enumerate}
\item \emph{Imperfect sensing:} Insufficient resolution of the radar function may cause an estimation error resulting in inaccurate beam alignment \cite{10633859}.
\item \emph{Association timeout:} Frequent beam switches induce a long time delay that is intolerable for
latency-sensitive services, referred to as association timeout \cite{10633859}.
\item \emph{Blockage:} The SSB transmission is blocked by nodes, objects, or scatterers \cite{9765510}. 
\item \emph{Mobility:} The MT moves too fast and the scheduled SSB has not been received before reaching the beam or cell boundaries \cite{9765510}. 
\end{enumerate}
Considering imperfect sensing and association timeout, the beam misalignment probability can be defined as 
\begin{equation}\label{bm1}
    P_{\mathrm{bm}} = P_{\mathrm{err}} + P_{\mathrm{to}},
\end{equation}
where $P_{\mathrm{err}}$ denotes the probability of imperfect sensing and $P_{\mathrm{to}}$ denotes the probability of association timeout in the ISAC network. The definition presented in \eqref{bm1} has been adopted in \cite{10633859}. 

Now, assume that mobility- and blockage-related issues lead to beam misalignment. Assuming that no sensing information can be acquired, the beam misalignment probability is defined as 
\begin{equation}\label{bm2}
    P_{\mathrm{bm}}^{\mathrm{no-sen}} = P_{\mathrm{mm}}^{\mathrm{no-sen}}P_{\mathrm{ub}} + (1-P_{\mathrm{ub}}),
\end{equation}
where $P_{\mathrm{mm}}$ denotes the mobility-induced misalignment probability and $P_{\mathrm{ub}}$ denotes the probability of unblocked LoS transmission of SSB. Assuming that assistance from sensing information is available, the beam misalignment probability is defined as 
\begin{equation}\label{bm3}
    P_{\mathrm{bm}} = P_{\mathrm{rm}}P_{\mathrm{ub}},
\end{equation}
where $P_{\mathrm{rm}}$ denotes the resolution-related beam misalignment probability. The definitions presented in \eqref{bm2} and \eqref{bm3} have been used in \cite{9765510}. 

\end{itemize}

The main performance metrics presented in this subsection and their relationship to key performance indicators are summarized in Fig. \ref{fig:metrics}.

\begin{table*}[t!]
    \centering
    \caption{Presentation of the main system models investigated}
    \begin{tabular}{|p{4.6cm}|p{7.0cm}|p{4.2cm}|p{0.7cm}|}
        \hline
\textbf{System Model} & \textbf{Main Contributions} & \textbf{Time/Bandwidth Allocation} & \textbf{Refs} \\ \hline 
           BS, users, and targets follow PPP in 2D  & 1ooperative ISAC, coordinated beamforming, cooperative sensing\newline 
2: Cooperative sensing, cooperative communications, CRLB investigation, cluster size optimization in conjunction with power allocation\newline
3: The impact of network deployment density is studied on the JCAS SIR performance
 & 1,2: Same time and frequency blocks for communication and sensing\newline 
3: Communication and sensing occur in separate time-frequency resources
 & \cite{10735119}\newline\cite{10694260}\newline\cite{10769538}\newline \cite{10615428}\\
    \hline
    BS and users follow PPP in 2D, while targets are UAVs distributed in a circle at height h  & Maximize the energy efficiency & All BSs share the same transmission bandwidth & \cite{10490156} \\
    \hline Users and undesirable scatterers follow independent PPP, BS or AP position is fixed
    & 1: MmWave system with beamwidth investigations, in which antenna misalignment is studied\newline
2: Explore/exploit duty cycle, the transmitted power, (bistatic) radar bandwidth and pulse repetition interval for maximizing the network throughput\newline
3: MmWave system that takes into account the impact of clutter, antenna models RCS and interference
 & 1,3: Sensing and communication are carried out  simultaneously in different frequency bands\newline
2: Time division approach is employed
 & \cite{10433485}\newline \cite{ram2022optimization}\newline\cite{10278626} \\
    \hline
    BS and users follow independent PPP & 1: Evaluating the ergodic sensing rate constrained by the maximum communication rate, and the ergodic communication rate constrained by the maximum sensing rate\newline
2: Success serving probability is defined for both the communication and the sensing tasks based on mutual information
 & 1: Communication and sensing occur in separate time-frequency resources\newline
2: Sensing and communication are carried out  simultaneously in different frequency bands
& \cite{10556618}\newline \cite{10683162} \newline \cite{10622535} \\
    \hline
    Sensing infrastructure and fusion centers are modeled as Matern cluster process (FS are parent process following PPP and SI are the daughter process uniformly distributed around FC)  & Sectored antenna are assumed, delay upper bound of the sensory data transmission & Time division approach is employed & \cite{10412651} \\
    \hline
    BS follow PPP, blockage follow Boolean line process, users and sensing targets are mutually independent point processes (located at the center of the origin) & Multi-carrier waveform is assumed, shared waveform, directional beamforming, and monostatic sensing & Communication and sensing occur in separate time-frequency resources & \cite{10320397} \\
    \hline
    BSs follow PPP, while sensing targets and users distances from BS have been evaluated using a probabilistic approach & The distance-dependent blockage model is adopted, which integrates the effects of LoS, NLoS, and target reflection cascading paths & Time division approach is employed & \cite{10579074} \\
    \hline
     BSs, mobile terminals and blockers follow PPP & 1: MT are moving in a straight line, and beam misalignment is investigated\newline
2: THz, MT are moving in a straight line, minimization of beam misalignment is performed, coverage probability is investigated
 &Orthogonal resources in time and frequency

& \cite{9765510}\newline \cite{10437749}\newline \cite{10633859} \\
    \hline
   Macro BS and small BSs follow independent PPP, UEs and targets also follow independent PPPs  & RIS-aided NOMA scheme is investigated, interference comes from other BSs & Time division approach is employed  & \cite{10565853} \\
    \hline
    Primary and secondary network transmitters follow independent PPP, each blockage follows an independent stationary point process & In a cognitive network, a sense and predict approach is used to maximize the ASE of secondary networks & Time division approach is employed  & \cite{8681732} \\
    \hline
    BSs are distributed according to $\beta$-GPP, the points of interest are distributed according to PPP & Full duplex, self interference between radar and communication antennas, a cooperative multi-point radar detection technique is proposed & Time division approach is employed  & \cite{9420372} \\
    \hline
    APs are distributed as PPP and tagged user and sensing targets are assumed at the origin & THz communication study, in which the human body blockage has been investigated & Three waveform design schemes are compared frequency-division ISAC, time-division ISAC,  and fully-unified ISAC & \cite{10622748} \\

        \hline
  APs are distributed according to PPP, UEs are uniformly distributed on a circle around the transmitters  & CSMA/CA-based channel access mechanism is used, the sensing performance is analyzed in terms of the maximum unambiguous range & Time division approach is employed   & \cite{10571293} \\
    \hline
    
    BSs and radars are distributed according to independent PPPs, users are distributed uniformly and independently & Dynamic transmission strategy is proposed & Resource blocks are assigned in both communication and sensing & \cite{10436875} \\

        \hline
    
    BSs are distributed according to the PPP communication mode UEs are distributed uniformly and independently at random in the Voronoi cells of the BSs & Successive interference cancellation is proposed & Resource blocks are assigned in both communication and sensing & \cite{10570658} \\
    \hline

        Road infrastructure, vehicles, and obstacles follow a superimposed generalized PPP in 2D & Cooperative JCS cooperative detection model, calculate the probability of successful JCS detection
and communication
 & Same time and frequency blocks for communication and sensing & \cite{9931437} \\

     \hline

        Vehicles and roadside units follow independent PPPs in a two-lane vehicular network & 1: ISAC performance evaluation in SG, monte-carlo, and ray-tracing frameworks \newline
2: Introduction of joint coverage and detection performance metrics for ISAC, cooperative detection with interference cancellation
 & Same time and frequency resources for sensing and communication & \cite{de2024full}\newline \cite{10569084}\\

      \hline

    \end{tabular}
    \label{tab:system_model1}
\end{table*}

\setcounter{table}{1} 
\begin{table*}[h]
    \caption{Presentation of the main system models investigated (continued from previous page).}
    \centering
    \begin{tabular}{|p{4.6cm}|p{7.0cm}|p{4.2cm}|p{0.7cm}|}
        \hline
\textbf{System Model} & \textbf{Main Contributions} & \textbf{Time/Bandwidth Allocation} & \textbf{Refs} \\ \hline

        BSs and UEs follow independent PPPs in 2D & Joint consideration of power and spectrum allocation for ISAC, design of an adaptive beamwidth strategy for communication
 & Time division or frequency division approach is employed & \cite{10562219} \\

       \hline

        BSs, UEs, and UAVs follow independent HPPPs in 2D & Cooperative ISAC for UAV surveillance, optimization of network configurations for sensing and communication performance
 & Same time and frequency resources for communication and sensing & \cite{10747175} \\

        \hline

        BSs follow cellular structure in 2D, UAVs follow independent PPPs in 2D and there are uniformly distributed buildings (obstacles) & Co-design of communication and bistatic radar sensing for UAV network resilience; novel radar cross-section estimation method
 & Non-overlapping spectrum allocation between communication and sensing & \cite{9893396} \\

         \hline

        UAV with radars and UAV for communications follow independent 2D HPPPs & Closed-form expressions for successful ranging probability and transmission capacity
 & Spectrum overlay multiple access and time-division multiple access
 & \cite{10217344} \\

          \hline

        Vehicles and roadside units follow a Cox PPP & Joint communication and computation-assisted sensing analysis with age of information as a key performance metric
 & Same time and frequency resources for sensing and communication
 & \cite{9814629} \\

           \hline

        Vehicles follow a 1D Mat\'ern hard-core process of type II & Design of a dual-beam ISAC scheme enabling $360^o$ radar detection and directional communication
 & Same time and frequency resources for sensing and communication
 & \cite{10682039} \\



             \hline

        BSs, users and sensing targets follow independent HPPPs in 2D & Design of cooperative beamforming to mitigate interference
 & Same time and frequency resources for sensing and communication
 & \cite{10695929} \\

              \hline

        Sensing targets follow a spatial HPPP in 2D & 1: UAV trajectory optimization for ISAC with SAR, considering energy efficiency\newline
2: UAV trajectory optimization for communication-assisted radar sensing
 & Same time and frequency resources for sensing and communication
 & \cite{10376044}\newline \cite{9847217} \\

               \hline

        Vehicles follow a 1D PPP  & 1: Derivation of closed-form expressions for cooperative detection range\newline 
        2: A CAS scenario is investigated for automotive radars
 & 1: Shared spectrum allocation between radar and communication\newline 2: Frequency division approach is employed. 
 & \cite{9606367}\newline \cite{huang2019v2x} \\
    \hline
UAVs, UEs, targets, and blockages follow independent HPPPs in 2D & Analytical expressions for the communication coverage probability and
successful detection probability are provided under the impact of blockage & Not specified & \cite{liu2024impacts} \\ \hline
    
    \end{tabular}

\end{table*}
\section{System Models}
The scope of this section is to present the main system models investigated in Sections IV and V, accompanied by a brief description of their technical characteristics. To this end, two tables have been prepared, namely Table~\ref{tab:system_model1} and Table~\ref{tab:system_model2}. To understand the main objectives and novelties of the research performed in the area of sensing and communication systems analyzed by SG, in Table~\ref{tab:system_model1}, an attempt is made to categorize the system models studied. A general observation that emerges from this table is that three main types of networks have been considered, namely terrestrial, aerial, and vehicular. However, for each of these types, a variety of different system models have been investigated, in which the SG principles have been employed to model the positions of BSs and/or users and/or targets and/or blockages and/or scatterers. Some further comments related to this table are the following:
\begin{itemize}
    \item 3D point process models are very rarely used, even in aerial communications scenarios;
    \item The general scenario where the positions of all network elements is modeled using SG has not been studied;
    \item In most cases, PPP has been employed, usually to simplify the analysis;
    \item In several studies, the main contribution was related to beamforming (coordinated or not) issues, e.g., beam alignment, interference cancellation (or mitigation);
    \item In many cases, the same time and frequency resources have been used for both sensing and communication;
    \item There are cases, where time (or frequency) division approach is employed;
    \item A very limited number of studies focus on CAS scenarios; 
    \item The impact of large-scale fading, i.e., shadowing, is not taken into account;
\end{itemize}

Key insights can also be derived from Table~\ref{tab:system_model2}, which can be summarized as follows.
\begin{itemize}
    \item An important part of the research has omitted antenna patterns from the analysis;
    \item The impact of interference has been considered in the presented analysis;
    \item In some cases, the effects of blockage and/or clutter have been investigated;
    \item Several studies have ignored propagation conditions with respect to the presence or absence of a LoS component;
    \item Small-scale fading has been taken into account is most cases, at least for the communication functionality;
    \item Multiple antenna studies have been reported, but their number is relatively limited;
    \item In most studies, the presented approach is multiple access scheme agnostic;
\end{itemize}
In terms of system models, two main categories have been identified, namely terrestrial and vehicular/aerial networks. Based on these categories, the reviewed works are summarized in the following sections.

\begin{table*}[t]
    \centering
    \caption{System model technical components (``non FT" stands for non Flat-Top antenna pattern).}
    \label{tab:system_model2}
    \renewcommand{\arraystretch}{1.2}
    \begin{tabular}{|p{1.1cm}|p{0.8cm}|p{1.4cm}|p{1.5cm}|p{2.0cm}|p{1.2cm}|p{1.4cm}|p{1.9cm}|p{1.9cm}|}
        \hline
        \textbf{Reference} & \textbf{HPPP} & \textbf{Antenna Pattern} & \textbf{Interference} & \textbf{Blockage/Clutter} & \textbf{LoS/NLoS} & \textbf{Small-scale Fading} & \textbf{MISO / SIMO / MIMO} & \textbf{Multiple Access Scheme} \\
        \hline
        \cite{10556618} & \checkmark & X & \checkmark & Blockage & X & \checkmark(c) & X & X \\
        \hline
        \cite{10735119} & \checkmark & X & \checkmark & Clutter & X & \checkmark(c) & MISO & X \\
        \hline
        \cite{10490156} & \checkmark & X & \checkmark & Clutter & X & \checkmark(c,s) & MISO & X \\
        \hline
        \cite{10412651} & X & \checkmark & \checkmark & X & X & \checkmark(c,s) & X & X \\
        \hline
        \cite{10769538} & \checkmark & X & \checkmark & X & X & \checkmark(c) & MISO & X \\
        \hline
        \cite{10579074} & \checkmark & X & \checkmark & Blockage & \checkmark & \checkmark(c,s) & X & X \\
        \hline
        \cite{meng2024network} & \checkmark & X & \checkmark & X & X & \checkmark(c) & MISO & CoMP-CDM \\
        \hline
        \cite{10565853} & \checkmark & X & \checkmark & X & \checkmark & \checkmark(c,s) & MIMO & NOMA \\
        \hline
        \cite{10217179} & \checkmark & X & \checkmark & X & X & \checkmark(c,s) & X & X \\
        \hline
        \cite{ram2022optimization} & \checkmark & \checkmark & X & Clutter & X & X & X & X \\
        \hline
        \cite{10320397} & \checkmark & \checkmark & \checkmark & Clutter & \checkmark & \checkmark(c,s) & X & X \\
        \hline
        \cite{9420372} & X & \checkmark & \checkmark & X & \checkmark & \checkmark(c,s) & X & X \\
        \hline
        \cite{10433485} & \checkmark & \checkmark (non FT) & \checkmark & Clutter & \checkmark & \checkmark(c,s) & X & X \\
        \hline
        \cite{10568512} & \checkmark & \checkmark & \checkmark & X & X & \checkmark & X & X \\
        \hline
        \cite{8681732} & \checkmark & X & \checkmark & Blockage & X & \checkmark & X & X \\
        \hline
        \cite{10633859} & \checkmark & \checkmark (non FT) & \checkmark & Blockage & X & X & X & OFDM \\
        \hline
        \cite{9765510} & \checkmark & X & X & Blockage & X & X & X & X \\
        \hline
        \cite{10569084} & \checkmark & X & \checkmark & X & X & \checkmark(c,s) & X & X \\
        \hline
        \cite{10582835} & X & \checkmark (non FT) & \checkmark & X & X & \checkmark(c,s) & MIMO & OFDM \\
        \hline
        \cite{9931437} & X & X & \checkmark & X & \checkmark & \checkmark(c,s) & X & X \\
        \hline
        \cite{9893396} & \checkmark & \checkmark (non FT) & X & X & X & X & X & CDM \\
        \hline
        \cite{10217344} & X & X & \checkmark & X & X & \checkmark(c,s) & X & SOMA/TDMA \\
        \hline
        \cite{10562219} & \checkmark & X & \checkmark & Clutter & X & \checkmark(c,s) & X & TDMA/FDMA \\
        \hline
        \cite{10747175} & \checkmark & \checkmark & \checkmark & Clutter & \checkmark & \checkmark(c,s) & MISO & X \\
        \hline
    \end{tabular}
\end{table*}

\section{Terrestrial Networks}
In order to better understand the commonalities and differences among the reviewed works, the contributions have been classified into ISAC, JCAS, and SAC systems. 

\subsection{Integrated Sensing and Communication}
In \cite{10556618}, a generalized SG framework was introduced to model ISAC networks, focusing on coverage and ergodic rate performance. BSs and mobile users (MUs) were distributed as HPPPs. The framework captures the spatial randomness in multi-cell networks and evaluates the coupling effects in the two functionalities, i.e., sensing and communication. Performance metrics included the coverage rate and the ergodic rate for both sensing and communication functions. Denser networks were shown to significantly improve ISAC coverage, with an increase in BS density from 1 km$^2$ to 10 km$^2$ improving coverage from 1.4\% to 39.8\%. Furthermore, the study revealed that increasing the constrained sensing rate improves the communication rate, although the reverse effect is less pronounced. 

In \cite{10735119},  a cooperative ISAC network was proposed using coordinated beamforming for interference nulling. The distribution of BSs, users, and targets was modeled as a PPP. The performance of the considered scheme was evaluated using the ASE criteria for both sensing and communication operations. Interference nullification was shown to improve the average data and radar information rates. Such an approach is necessary to maximize the sensing ASE, but not always for the communication ASE. The fixed relationship between the number of service users and the BS transmit antennas was also revealed for optimal performance. 

In \cite{10490156}, an analytical framework was introduced for analyzing sensing and communication performance in ISAC networks. The locations of the BSs, users, and targets were assumed to be modeled as PPPs. Closed-form expressions were obtained for evaluating PSE and EE. Moreover, based on the derived results, an optimization problem for optimal BS density was formulated and validated through Monte Carlo simulations. It was revealed that the optimal BS density for ISAC networks is lower than that for communication-only networks, improving coverage radius and energy efficiency, especially in low transmit power regions.

In \cite{10622535}, the theoretical performance of ISAC networks was studied, where the BSs and UEs were modeled as independent HPPPs. Each UE accesses the nearest BS to form a Voronoi tessellation, and each BS has its corresponding sensing target. Each target is randomly located in a disk region with the corresponding BS as the center. The paper introduced the success serving probability as a unified analytical framework based on mutual information for both communication and sensing. The study highlighted the importance of beamforming and the distribution of sensing targets in optimizing ISAC network performance.

In \cite{10412651}, the Mat\'ern cluster process (MCP) was used to model the spatial locations of the sensing infrastructures and fusion centers in an ISAC network. Note that the fusion centers are collection points for sensing information. Detection and transmission occur on the same millimeter wave (mmWave) spectrum, with orthogonal sub-bands allocated to SIs to reduce intra-cluster interference. The service capability of sensory data transmission is analyzed, which is characterized by the successful sensing probability. In addition, the delay upper bound was investigated using stochastic network calculus, which was subsequently transformed into an SNR-based metric. An optimal power allocation strategy was proposed to minimize the transmission delay, balancing the sensing quality and latency. 

In \cite{10683162}, the fundamental limits of the performance of ISAC were evaluated under resource constraints. A generalized SG framework was proposed in which the location of BSs and users are modeled using an HPPP. BSs perform positioning and communication functions during downlink transmissions, while users access positioning pilots from the nearest BSs for self-positioning while receiving communication data from the nearest BS. Theoretical results related to the coverage rate highlighted the impact of the BS density on ISAC performance, showing an improvement in the coverage rate from 1. 4\% to 39. 8\% as the BS density increases from 1 km$^2$ to 10 km$^2$. 

In \cite{10694260}, a cooperative ISAC scheme was proposed that integrates CoMP joint transmission and multi-static sensing. By optimizing the size of the cooperative BS cluster, the performance of sensing and communication (S\&C) performance was balanced. The network model included BSs, users, and targets, all modeled as independent PPPs. The authors quantified ISAC performance in terms of data rate and Cramér-Rao lower bound (CRLB), applying SG techniques. The simulation results indicated that the cooperative scheme outperforms the time-sharing scheme with more resources and a larger backhaul capacity. 

In \cite{10769538}, ISAC performance was quantified in terms of data rate and CRLB, applying SG techniques for analysis. In the network considered, each BS was assumed to support multiple transmit and receive antennas, with locations modeled as an HPPP. Communication users and targets were also modeled as independent PPPs. By optimizing cooperative cluster sizes and power allocation, the data rate-CRLB region and the weighted sum of data rate and inverse CRLB were maximized. The CRLB expression showed that the deployment of $(N)$ ISAC transceivers enhances the cooperative sensing performance according to the scaling law $(\ln^2 N)$.  The simulation results presented showed that the proposed scheme improves the data rate and reduces the CRLB, outperforming conventional time-sharing schemes.

In \cite{10579074}, the coverage performance of ISAC networks was examined under blockage effects. In this context, a general analytical model was proposed that utilizes time-division duplexing to avoid mutual interference. The locations of BSs were modeled as PPP, while for simplicity, sensing targets were considered as point scatterers, primarily detectable under line-of-sight (LoS) conditions. The performance metrics used included the communication SINR and sensing coverage probability. It was shown that increasing the BS density improves the coverage probability when the BS density is lower than the blocking density. Moreover, an optimal BS density was identified for maximizing coverage probability, with blocking providing favorable gains by reducing interference.

In \cite{10695883}, the gain in the coverage rate from reconfigurable intelligent surfaces (RISs) in ISAC networks was theoretically investigated. The BSs, equipped with multiple antennas, and blockages were modeled as PPP. RISs were deployed to enhance ISAC performance, with optimal densities of BSs and RISs identified for practical deployment. Using mmWave beamforming, user association policies, and conditional coverage rates were derived, with the marginal coverage rate calculated using distance-dependent thinning. It was shown that a joint coverage rate improvement from 67.1\% to 92.2\% is achieved with RIS deployment. 

In \cite{10636567}, a cooperative ISAC scheme was investigated, revealing critical cooperative dependencies in the ISAC network. Specifically, a coordinated beamforming approach was proposed for adaptive interference nulling, with each BS transmitting to multiple users and performing sensing operations. The study derived the expression for the sensing and communication ASE to analyze spatial resource allocation objectives, including spatial diversity gain, spatial multiplexing gain, and interference nulling. The BS, users, and target locations followed an HPPP. It was shown that interference nulling is not necessary for maximizing communication ASE but is required to maximize sensing ASE.  

In \cite{10622748}, a framework for THz ISAC networks was presented, which also considered the blockage effects, antenna radiation, and association schemes. The system model involved random access point deployment via a 2D HPPP, with APs serving as transmitters and transceivers for communication and sensing. Analytical expressions for the CP and sensing probability were derived using the moment generating functions (MGFs) of aggregated interference. Network capacity analysis compared three THz ISAC waveform designs, showing that high AP density maximizes communication and sensing coverage probabilities. The narrower beamwidth improves the average communication rate, while the average sensing rate initially increases and then decreases.

In \cite{10570658}, a large full-duplex ISAC-enabled cellular network was studied to efficiently reuse the scarce spectrum. By modeling the spatial locations of the BSs and UEs as independent PPPs, the detection probability at the BS was investigated based on the received SINR metric. To this end, both self-interference cancellation and successive interference cancellation schemes were utilized to successfully decode communication and sensing signals at the BS. Numerical results revealed that after a certain level of residual self-interference, decoding/detection is not viable, while the sensitivity of both radar and communication modes were demonstrated.

In \cite{meng2024network}, a cooperative architecture was proposed for ISAC-enabled networks, incorporating coordinated multipoint transmission (CoMP) along with multistatic sensing. To account for practical aspects, the impact of the allocation of antennas to base stations on cooperative sensing and cooperative communication performance was investigated. By modeling the spatial locations of the BSs as a 2D PPP, analytical expressions for the communication data rate and the CRLB were obtained. Subsequently, the paper focused on the sensing performance by investigating three localization techniques to assess their effects on ISAC network performance. After investigating the key performance metrics, a performance boundary optimization problem was studied. The results verified that cooperative transmission and sensing in ISAC networks can effectively improve the sensing and communication gains and strike a more flexible trade-off between sensing and communication performance. In addition, the proposed cooperative scheme shows superior performance improvement compared to centralized or distributed antenna allocation strategies, especially when using a large number of antennas.   

In \cite{10681925}, a digital twin (DT)-based model drift-adaptive resource reservation scheme for ISAC-enabled networks was proposed. The term model drift was defined as the degradation of the accuracy of a predictive model when the data distribution changes. Using tools from SG, two spatial models were adopted for the locations of ISAC devices and sensing targets. In the first model, the spatial locations of the ISAC devices and the sensing targets were modeled as independent 2D PPPs, while in the second model, the spatial locations of the ISAC devices and the sensing targets were modeled as PPP and TCP, respectively. The goal was to provide a statistical quality of service guarantee with minimal resource consumption for a sensing service. To this end, a DT was constructed at the network controller to collect the locations of sensing nodes and targets. The simulation results showed that the proposed resource reservation scheme was able to reduce the resource under- and over-provision under non-stationary network conditions. 

In \cite{9538929}, the age of information (AoI) in massive mobile sensing networks was investigated by considering channel fading, path loss, and ISAC-enabled network topology, as well as the problem of shared spectrum interference. By modeling the spatial locations of the receivers and the sensing nodes as independent PPPs,  the upper and lower bounds of AoI were derived, and the relationship between AoI and mobile speed, spatial density, and channel access probability was demonstrated through simulations. The numerical results demonstrated that the average AoI decreases as the average speed of the interferers increases under a last-come-first-served discipline.

In \cite{10436875}, a dynamic transmission strategy was proposed in which the quantity of radar mode detection is traded for quality as a countermeasure to the fact that interferers in radar mode in ISAC-enabled networks have direct links. By modeling the spatial locations of the BSs and the radars as independent PPPs, the performance of both the radar and communication modes with and without a dynamic transmission strategy was investigated and compared to the performance of a conventional radar-only network. The numerical results revealed that the dense deployment of low-cost passive radars offers superior detection for farther targets, especially compared to a radar-only network.

In \cite{10565853}, an analytical framework was introduced to evaluate downlink transmissions in MIMO IoT heterogeneous networks (HetNets) with ISAC capability. By modeling the spatial locations of the NOMA-enabled BSs and the UEs as independent PPPs and by further utilizing the RIS technology, approximate and asymptotic expressions for outage probability were derived for i) a scenario involving direct transmission from the BS to the typical blocked user, and ii) a scenario involving transmission via active RIS. In addition, the approximate expressions for the ergodic rates, the system throughput, and the beam pattern for the sensing performance were also derived. The results emphasized the benefits of the proposed active RIS-NOMA over conventional orthogonal multiple access HetNets and revealed that an increase in the number of the RIS elements significantly improves the proposed active RIS-NOMA outage performance. 

In \cite{10217179}, a distributed ISAC network was investigated where ISAC users were randomly distributed in a circular area using an HPPP. These ISAC users share a common bandwidth and transmit power, and simultaneously transmit communication and sensing signals that interfere with each other. Performance was evaluated in terms of coverage and detection probabilities, which were obtained in closed form. Extensive simulations confirmed the accuracy of the derived results, highlighting the fundamental performance limits and trade-offs in distributed ISAC networks.

\subsection{Joint Communication and Sensing}
In \cite{ram2022optimization}, a JCAS system was investigated, focusing on bistatic/passive radar deployments with directional beams to identify MUs and establish communication links. The distribution of the scatterers was modeled as an HPPP. Performance metrics include network throughput, radar detection efficiency, and joint radar and communication reliability. Based on SG tools, an optimization of the JCAS parameters, e.g., exploration/exploitation duty cycle, radar bandwidth, transmit power, and pulse repetition interval, was performed to maximize throughput. Monte Carlo simulations validated the theoretical results, highlighting the trade-offs between detection success and clutter density, achieving a peak JCAS reliability of 70\% for bistatic configurations.

In \cite{10603243}, a novel framework was introduced for analyzing the probability of success in JCAS systems using the DPP. The distribution of the BS and MUs was modeled with the DPP, which captures repulsion among nodes. The performance of the system was evaluated using the success probability, based on the received SINR. The proposed framework was shown to provide a rigorous closed-form analysis of success probability, providing insight into the trade-offs between sensing performance and communication reliability. 

In \cite{10320397}, a rigorous analytical framework was presented and used to characterize the joint performance of communication and parameter estimation JCAS networks. The BSs and sensing objects were modeled as stationary, ergodic point processes. Performance metrics included coverage probability and ergodic rate, derived using mutual information and Fisher information matrix. The CP and ergodic rate concepts were extended to radar settings using mutual information. In addition, methods were developed to obtain bounds on the Laplace transform of shot noise processes, while a new analog of Hölder's inequality was introduced. Closed-form bounds and approximations for JCAS coverage and rate were provided and validated through numerical case studies, highlighting the trade-offs between sensing and communication performance. It was shown that the sensing SINR improves with BS density, while interference impacts sensing more than communication. 

In \cite{10615428}, a fine-grained analytical framework for JCAS networks was proposed by deriving the meta distribution of the SIR. The locations of BSs, UEs, and sensing objects were modeled as independent HPPPs. The performance of the system was evaluated in terms of the conditional JCAS coverage probability and the complementary cumulative distribution function of the SIR meta distribution. An important outcome of the paper was the investigation of the impact of interference from simultaneous transmissions.

In \cite{10571293}, the performance of a JCAS network where nodes switch between radar and communication modes in an uncoordinated manner was investigated. The nodes followed a CS-based medium access protocol, and their locations are modeled as a PPP. Key contributions included the development of an analytical framework for evaluating CSMA-based JCAS networks, the modeling of interference from uncoordinated transmissions, and the analysis of sensing performance in terms of maximum unambiguous range. The study also derived closed-form expressions for mean access probability and aggregated network throughput density, which were validated through extensive simulations. These findings help to characterize both radar and communication performance in large-scale JCAS networks.

In \cite{8855935}, the performance of JCAS systems for multi-radar cooperative detection in drone surveillance was investigated. Radars were deployed on a 2D plane to detect drones in a 3D space, forming clusters with a fusion center. The distribution of fusion centers followed an HPPP, while other radars follow an independent PPP. The performance of the considered scheme was studied in terms of the detection volume and successful communication probability. An ptimal power allocation between radar and communication was shown to maximize the detection volume. 

In \cite{9049687}, the performance of a time-sharing JCAS network was studied, where the nodes (distributed as HPPP) switch between radar sensing and data transmission modes. The study proposed approximation methods to evaluate various performance metrics, such as the maximum detectable radar range under a desired false alarm probability, packet loss rate, and per-node throughput, offering insights into resource allocation strategies, such as communication packet length and radar update rate.  The trade-offs between radar and communication performance in a shared bandwidth scenario were highlighted.

In \cite{9420372}, the co-existence of full-duplex JCAS systems in mmWave HetNets with repulsive node distributions was investigated. The study developed a mathematical framework using SG to model these networks, considering the spectrum sharing and radar sensing capabilities of the BSs. Specifically, the BS distribution was modeled as a $\beta$-GPP. In this context, a cooperative multi-point radar detection technique with three hard-decision combining rules (OR, Majority, AND) was proposed, and analytical expressions for detection performance were provided. Repulsive BS deployment was shown to improve detection performance by reducing interference, and temporal interference correlation significantly affects network performance.

In \cite{10278626}, an SG-based framework was used to analyze the performance of a mmWave JCAS system. The system included a sensing subsystem for radar detection of MUs and a communication subsystem using directional antennas. Both functions operated simultaneously to reduce delay and improve beam alignment. The network was modeled with MUs and scatterers whose distribution follow PPP, and JCAS nodes in hexagonal cells acting as monostatic radars and BSs. The performance of the system was evaluated in terms of CP and DP, with parameters of interest the MU and clutter density, RCS fluctuations, radar search duration, antenna directivity, and bandwidth. Radar sensing was shown to significantly affect overall performance, with the duration of radar search being critical to maximizing the average system throughput. 
 
\subsection{Sensing Assisted Communications}
In \cite{10433485}, an analytical model for mmWave radar-assisted communication systems was presented, considering undesirable clutter, RCS fluctuations, and interference from the transmitting nodes. MUs and clutter scatterers were distributed as a PPP. The network consisted of hexagonal small cells, each with a BS-mounted radar at the center, in which the beamwidth trade-offs were investigated. The system performance was evaluated using the system throughput and antenna misalignment error. Key results highlighted the importance of beamwidth design trade-offs, showing that careful selection can eliminate beam training overhead and improve system performance.

In \cite{10568512}, a mmWave cognitive network with primary and secondary links was investigated, using directional channel sensing and communication on the same frequency band. Secondary transmitters were assumed to be distributed as PPPs. The study developed an analytical framework showing that primary operators can restrict secondary transmissions based on signal power thresholds, creating asymmetric protection zones. The coverage probabilities were computed for primary and secondary links, highlighting the impact of various parameters on network performance. The independence of channel access probabilities of secondary transmitters from network density and the trade-off between antenna gain and beamwidth were demonstrated.

In \cite{8681732}, the interference mismatch problem was investigated in sensing-based underlay cognitive radio (CR) networks by exploiting spatial correlations between secondary transmitters and receivers. These correlations depend on the locations of the interferers (following a PPP), blockages, and directional beams. A spatial sensing framework was proposed, where the secondary transmitter access control is based on a single sensing result, assuming static behavior of the primary transmitter during the observation period. The interference prediction model used outage probability to represent spatial correlations, facilitating a probabilistic approach to access control and interference management. This framework based on outage probability was used to predict SIR levels in different scenarios. Moreover, ASE optimization was performed through linearly mapping access thresholds to decoding targets, balancing access probability and transmission success.

In \cite{10633859}, a joint synchronization signal block and reference signal-based sensing (JSRS) scheme was proposed to assist beam alignment in THz networks based on 5G beam management. By modeling the spatial locations of BSs, UEs, and blockers as independent PPPs, the analytical expression for communication coverage probability for the JSRS-enabled network was obtained. The expression was used for evaluating the network's ability to reduce beam misalignment and quantify the coverage improvement. The expression also guides the network density deployment and beamwidth selection. In terms of sensing operation, the time-to-frequency allocation ratio that minimizes the sensing error-induced beam misalignment was obtained by solving the formulated optimization problem. The numerical results demonstrated that the proposed JSRS scheme is effective and highly compatible with the 5G air interface.  


In \cite{10437749}, a time-frequency resource allocation technique for sensing signal mapping schemes that maximize the coverage probability of the ISAC-THz networks with reduced sensing costs was proposed. By modeling the spatial locations of the BSs, mobile terminals, and blockers as independent PPPs, the trade-off between coverage performance and sensing accuracy was demonstrated. To this end, a narrow beam design was adopted, and the optimal trade-off between sensing assistance and its cost was obtained as a resource allocation-based optimization problem. The results revealed that wider coverage requires more sensing resources, and the high frequency significantly limits the unambiguous sensing range.

In \cite{9765510}, a beam alignment scheme was proposed utilizing SSB-based sensing to assist beam switching. By modeling the spatial locations of the BSs, UEs, and blockers as independent PPPs and determining the optimal SSB pattern design, a closed-form resource allocation was derived that is independent of the density and mobility of the nodes. The numerical results revealed that the proposed scheme significantly reduces the beam misalignment probability in highway and urban scenarios. 

In \cite{10549451}, the performance of an ISAC-enabled network was investigated in terms of achievable throughput, where the nodes utilize the radar functionalities to direct highly directional beams to MUs for communication services. By modeling the spatial locations of the clutter scatterers as PPPs, radar scanning was performed to detect and localize MUs/targets in the presence of discrete clutter scatterers. Subsequently, radar information was utilized to analyze the radar operating metrics and  realize high communication throughput. The numerical results demonstrated the impact of radar duty cycle and beamwidth on the performance of the ISAC-enabled network in terms of communication throughput. 

\subsection{Comparative Observations}
By comparing the novel advances presented in this section, the following key observations have been identified. Despite the fact that a very large number of works have integrated both optimization and SG tools into the ISAC paradigm \cite{10735119,10695883,10769538,10437749,9765510,10633859,ram2022optimization,meng2024network}, the use of AI and ML tools remains remarkably unexplored. Moreover, although it is undoubtedly a key 6G enabling technology, the effect of RISs on the performance of ISAC networks has hardly been explored \cite{10695883,10565853}, while the key performance limitations have not been investigated at all. Finally, it is surprising to observe that despite the fact that with recent advances in ISAC technology, 3GPP-based mobility modeling has become more essential than ever, no research work has performed a performance comparison of different mobility schemes, similar to \cite{9078878}.

\section{Aerial/Vehicular Networks}
For the scenario of aerial or vehicular communication networks, the contributions are also classified in the ISAC, JCAS, and SAC scenarios. 
\subsection{Integrated Sensing and Communication}
In \cite{de2024full}, the performance of an ISAC vehicle-to-everything (V2X) downlink scenario was evaluated, where a vehicle simultaneously detects the next vehicle ahead while receiving a communication signal from a roadside unit (RSU). The system model considered RSUs distributed using PPP, while key performance metrics included coverage probability, detection probability, and false alarm rate. The paper analyzed univariate and joint radar and communication metrics using SG (SG), Monte-Carlo, and ray-tracing (RT) frameworks. System parameters were extracted from ray-tracing simulations, and the metrics were compared to evaluate the accuracy of the SG framework. The results showed that the SG and Monte Carlo models were relevant with respect to ray-tracing simulations for univariate metrics, although larger discrepancies were observed for joint metrics.

In \cite{10569084}, a full-duplex ISAC downlink system was investigated for automotive scenarios. The system model considered a vehicular network with vehicles traveling in the direction of the typical vehicle, interfering vehicles traveling in the opposite direction, and RSUs distributed on three parallel lines following three independent PPPs. Key performance metrics included coverage probability, false alarm probability, and detection probability. The framework was validated through Monte Carlo simulations, which highlighted trade-offs between sensing and communication functions. The work explored the optimization of network node performance and density to maximize joint performance. It also extended the analysis to account for mutual interference between sensing and communication, considering both perfect and imperfect interference cancellation.

In \cite{10682039}, a dual-beam ISAC scheme was proposed for vehicular networks to provide 360° radar detection and directional communication simultaneously. The study developed a SG-based analytical framework using the MHCP to model the vehicle positions and evaluate the detection probability and communication coverage probability, considering both incident and reflected interference. It investigated two power allocation optimization problems to improve detection performance without degrading communication performance. Simulation results confirmed the accuracy of the analytical models and demonstrated the benefits of the proposed power allocation schemes in improving the average detection probability.


In \cite{10695929}, the coverage performance of air-ground ISAC downlink networks was analyzed. Specifically, the coverage probability for communication was evaluated while taking into account interference during sensing. The work incorporated cooperative beamforming to improve sensing accuracy and compared HPPP models with lattice models to establish lower and upper performance bounds.

In \cite{10376044}, ISAC was studied for UAV-borne synthetic aperture radar (SAR) systems, focusing on minimizing power consumption while maintaining communication and sensing performance. It considered an aerial network where sensing targets were distributed using a PPP. A trajectory optimization framework was proposed to reduce propulsion power under sensing and communication constraints, with emphasis on power consumption and average rate.

In \cite{10582835}, inter-vehicle communication scheme to schedule communication and sensing signals for vehicles was investigated. By exploiting the 1D-MHCP for modeling the spatial locations of vehicles, an analytical framework to statistically characterize the mutual interference among multiple vehicles was proposed. The interference mitigation scheme was evaluated in terms of interference probability, duration, and the achievable detectable density. Going one step further and recognizing the different performance requirements of communication and sensing functions, a joint resource allocation problem was investigated. Subsequently, the aforementioned optimization problem was solved by exploiting an algorithmic approach. The results demonstrate a notable enhancement in the proposed ISAC-based interference mitigation scheme as compared to benchmarks.

In \cite{liu2024impacts}, the communication coverage and successful detection probabilities was investigated in UAV-enabled ISAC systems. The main novelty of the paper rises on the fact that the impact of blockages on propagation paths was considered for deriving probability distributions for UAV-user and UAV-target distances. PPP were used to model the 2D dimensions of the UAVs, users, targets and blockages. The analytical framework presented was used to examine how UAV density, height, and blockage height influence system performance, demonstrating that strategic UAV deployment can effectively compensate for blockage-induced degradation.

\subsection{Joint Communication and Sensing}
In \cite{9931437}, the performance of cooperative detection in JCAS vehicular networks was investigated. Using geographic information system data, road obstacles were statistically modeled and the effects of shading and SG obstacles were evaluated using a generalized PPP. Key performance metrics included the probability of successful cooperative detection under LoS and NLoS conditions. The paper also explored simulation-based deployment optimization strategies to improve detection reliability and communication efficiency, and emphasized adherence to urban roadway structure standards to ensure practical relevance. Finally, it focused on obstacle detection as a primary service in the downlink and demonstrated the potential of JCAS to enable coordinated detection and infrastructure optimization in urban vehicular networks.  
\begin{table*}[t!]
    \centering
     \caption{Summary of the key results of the reviewed literature.}
        \renewcommand{\arraystretch}{1.3}
    \begin{tabular}{|p{2.5cm}|p{1.5cm}|p{10.5cm}|p{1.5cm}|}
    \hline
    \textbf{Area of Interest} & \textbf{Network Type} & \textbf{Key Result} & \textbf{Reference}
 \\ \hline 
      Spatial Resources & Terrestrial & 1) The optimal communication trade-off for ASE maximization tends to use all spatial resources for multiplexing and diversity gain without interference nulling. \newline
    2) For sensing objectives, sufficient antenna resource allocation tends to eliminate inter-cell interference. & \cite{10735119} \\
    \hline
    BS Density & Terrestrial & 1) Increasing the BS density from 1 km$^{-2}$ to 10 km$^{-2}$ can increase the ISAC coverage rate from 1.4\% to 39.8\%. \newline
    2) The sensing function performs best at high BS densities, while the communication performance is optimized at moderate BS densities. \newline
    3) Even at low BS densities, interference remains a more significant factor in the performance of sensing than in communication. \newline
    4) With the increase of the constrained sensing rate, the ergodic communication rate improves significantly, but the reverse is not obvious. \newline
    5) The small distance resolution and the high data rate are the two main constraints on the maximum achievable coverage probability with optimized power and spectrum allocations. \newline
    6) We should not pursue resolution redundancy too much in practical applications of JSAC, since maintaining redundancy in sensing will cripple the joint performance of JSAC. \newline
    7) If we want to improve the maximum coverage probability by increasing the small distance resolution in sensing or decreasing the high data rate in communication, the latter option may be a better choice. \newline
    8) If there is some information about the target density in practical applications, the corresponding optimal BS density can be set. If there is no information about the target density in practical application, the density of BS can be set as 150 to 200 points/km$^2$. \newline
    9) Repulsive BS deployment improves detection performance by reducing interference. & \cite{10556618,10320397,10562219,9420372} \\
    \cline{2-4}
    & Terrestrial with Aerial Targets & 1) The optimal dual-function radar-communication BS density for the ISAC network is different from the classical communication-only network, and this optimal value becomes smaller as the target altitude increases. \newline
    2) A few BSs are sufficient to achieve optimal performance at very high altitudes. & \cite{10490156} \\
    \hline
    Blockage Density &  & 1) Blockages can have a positive effect on coverage, especially in improving communication performance. \newline
    2) There is an optimal BS density when the blockage is of the same order of magnitude as the BS density,  maximizing communication or sensing coverage probability. & \cite{10579074} \\
    \hline
     RU Height and Density & Vehicular & 1) Optimal deployment height of infrastructure is 7 m. \newline
    2) The infrastructure deployment interval is 150 to 300 m according to the probability requirement of successful cooperative detection, which complies with the Chinese and European transport standards. & \cite{9931437} \\
    \hline
    Antenna Beamwidth & Terrestrial & 1) The beamwidth of the communication antenna exhibits a trade-off where a narrow beam provides higher gain and reduces interference, but increases the effect of misalignment error. \newline
    2) There exists an optimal search duration and a main lobe spread parameter of the communication antenna at the optimal search duration value. \newline
    3) While maintaining the same beam alignment quality, sensing-aided networks can use narrower beams to extend network coverage. & \cite{10433485,9765510} \\
    \hline
    Effect of RISs & Terrestrial & A joint coverage rate improvement is achieved with RIS deployment. & \cite{10695883} \\
    \hline
    Cooperative vs. Standalone Sensing & Terrestrial & 1) The use of $N$ ISAC transceivers results in an improved average cooperative sensing performance across the network, following the $\ln2N$ scaling law. \newline
    2) The scaling law is less pronounced compared to the $N^2$ performance improvement achieved when the transceivers are equidistant from the target. \newline
    3) Cooperative transmission and sensing in ISAC networks can effectively improve the sensing and communication gain. \newline
    4) The cooperative scheme shows superior performance improvement compared to centralized or distributed antenna allocation strategies, especially when more antennas are available. & \cite{10769538,meng2024network} \\
    \cline{2-4}
    & Aerial & 1) Cooperative sensing can achieve a large performance improvement over standalone sensing by utilizing the spatial diversity gain of multiple collaborating BSs. \newline
    2) Multiple illuminators provide additional robustness for detection. \newline
    3) The detection performance of multiple illuminators can be improved if the detection of the same target by different illuminators is independent. & \cite{10747175,9893396} \\
    \hline
    Multiple Access Technique & Aerial & 1) TDMA outperforms SOMA in terms of successful ranging probability. \newline
    2) SOMA outperforms TDMA in terms of transmission capacity. \newline
    3) SOMA can outperform TDMA in terms of both successful ranging probability and transmission capacity when the node density of active UAV-radars is larger than the node density of UAV-communications. & \cite{10217344} \\
    \hline
    Safety Critical Applications (Collision Avoidance) & Aerial & A distributed communication and sensing system can improve the resilience of UAV networks. & \cite{9893396} \\
    \hline
    Interference & Vehicular & ISAC-based schemes can effectively mitigate mutual interference among multiple vehicles. & \cite{10582835} \\
    \hline
    \end{tabular}
        \label{tab:key_results}
\end{table*}

In \cite{9893396}, a distributed communication and sensing system was co-designed to improve the resilience of UAV networks. The system model considered an aerial network with UAV locations modeled as a PPP, while system reliability and UAV network resilience were investigated. The study analyzed a JCAS framework incorporating bistatic radar functionality to improve collision avoidance. A novel radar cross-section estimation method was proposed to address the challenges of modeling complex-shaped objects. The work combined radio network planning and link budget estimation with Monte Carlo simulations to evaluate radar detection performance. SG methods were used to estimate collision probability without extensive flight-hour simulations. The numerical results confirmed robustness to radar cross-section pattern nulls, demonstrating the suitability of the system for safety-critical applications such as collision avoidance.

In \cite{10217344}, the coexistence of UAV radar and communication networks was analyzed under different multiple access protocols. It considered an aerial network operating in the mmWave band, with UAV-mounted BSs and radars modeled as PPPs. Spectrum overlay multiple access (SOMA) and time division multiple access (TDMA) protocols were studied, and closed-form expressions for performance metrics such as successful range probability and transmission capacity were derived. The analysis showed that SOMA outperforms TDMA in both successful ranging probability and transmission capacity when the density of active UAV radars exceeds that of UAV communications. By comparing multiple access schemes, the paper highlighted the trade-offs between radar and communication performance and provided insights into optimizing UAV network configurations for JCAS systems.

In \cite{9847217}, JCAS was studied for cellular-connected UAVs used as bistatic SAR platforms to sense randomly distributed objects using PPP. A trajectory planning algorithm was proposed to optimize UAV flight paths to minimize propulsion energy while ensuring the required sensing resolutions. Key performance metrics included power consumption, with simulations that demonstrated energy savings of up to 55\% compared to the shortest-path approaches.

In \cite{9606367}, JCAS systems were analyzed for cooperative detection in vehicular networks. The system modeled vehicle positions using a 1D PPP and focused on improving spectrum utilization and mitigating interference. The paper derived closed-form expressions for the average cooperative detection range under different scenarios with varying numbers of vehicles. It was demonstrated that JCAS systems improve spectrum sharing and detection efficiency through JCAS. Analytical results validated the proposed framework and highlighted its effectiveness in evaluating cooperative detection performance.

\subsection{Sensing Assisted Communications}
In \cite{10562219}, the performance of JCAS for large-scale networks was analyzed, focusing on downlink sensing-assisted beamforming and adaptive resource allocation. The system model considered a vehicular network operating at 77 GHz, where the locations of users and BSs were modeled as two independent PPPs. The paper evaluated the coverage probability as a function of the user and BS densities, the required high data rate and small distance resolution, and the allocation ratios of power and spectrum. The sensing system detected the environment and obtained user positions, followed by adaptive beamforming to improve communication efficiency. Using SG, interference in sensing and communication was modeled separately and joint coverage probability expressions were derived. The analysis highlighted the trade-off between small distance resolution and high data rate as key constraints to maximize coverage probability with optimized power and spectrum allocation. It also showed that different BS densities should be considered for different scenarios, and the relationship between the user and the BS densities could serve as a reference for practical applications.

In \cite{10747175}, the performance of downlink perceptive mobile networks was investigated for autonomous and cooperative UAV surveillance. The system model considered an aerial network operating in the mmWave band, where BSs, UAVs, and ground user equipment were modeled as a two-dimensional PPP. Key performance metrics included coverage probability and successful detection probability, while mutual interference and resource contention between sensing and communication functions were analyzed, providing a basis for network optimization. It proposed a cooperative sensing strategy that combined monostatic and bistatic sensing techniques to improve the reliability of UAV surveillance. The simulation results verified the framework and highlighted the advantages of cooperative sensing over individual approaches.

In \cite{9814629}, the freshness of sensing information in vehicular networks was analyzed, focusing on the joint scheduling of communication and computational resources. The system model considered an uplink vehicular network scenario in which RSUs equipped with computational resources processed data using a two-stage tandem queue. The study investigated the AoI and derived its average expression under SG, modeling vehicular networks as PPPs. Key performance metrics included coverage probability and expected data rate under transmission constraints. The analysis emphasized the trade-off between communication and computational capacities to optimize resource utilization while ensuring real-time information delivery. Simulations validated the proposed framework and demonstrated the relationship between resource capacity and information freshness.

In \cite{huang2019v2x}, in a communication-sensing scenario, the authors proposed a framework to take advantage of V2X communication to efficiently allocate spectrum resources. The spatial distribution of interfering radars was modeled as PPP and interference is evaluated for different radar types. Moreover, the minimum spectrum required for zero-interference was analyzed, and simulations confirmed that the proposed approach achieves near-zero interference while optimizing spectrum usage.

\subsection{Comparative Observations}
Comparing the various contributions in this section, the following observations have been identified. In V2X networks, SG has been employed to model the position of the vehicles, thus adopting 1D spatial models. Despite the fact that dynamic network conditions have been assumed, the impact of Doppler effect and/or outdated channel state information and/or beam misalignment have not been considered. In addition, blockage induced by other vehicles is also a parameter not included in the presented analysis. As far as aerial networks are concerned, all studies have focused on aerial-to-ground scenarios, i.e., not on air-to-air communications. Moreover, the network entities, e.g., aerial BSs or UAV-targets, are considered as spatially distributed in a 2D plane. In addition, shadowing phenomena due to random variations in mean signal levels are not considered. Finally, the distribution of random scatterers is not investigated for both vehicular and aerial investigations, while it is important to note that scenarios with combined UAV-supported V2X communication are not investigated.

\section{Key Results, Open Problems, and Future Directions}

This section presents a comprehensive summary of key findings and attempts to identify research challenges and future research directions for ISAC systems. First, key findings are presented for communication and sensing networks in the context of SG. Next, key research challenges and future directions are discussed in an effort to bridge the research gap between ISAC and SG towards the vision of a unified SG-based ISAC framework. 


  \begin{figure*}[t!]
  \centering
    \includegraphics[width=7in]{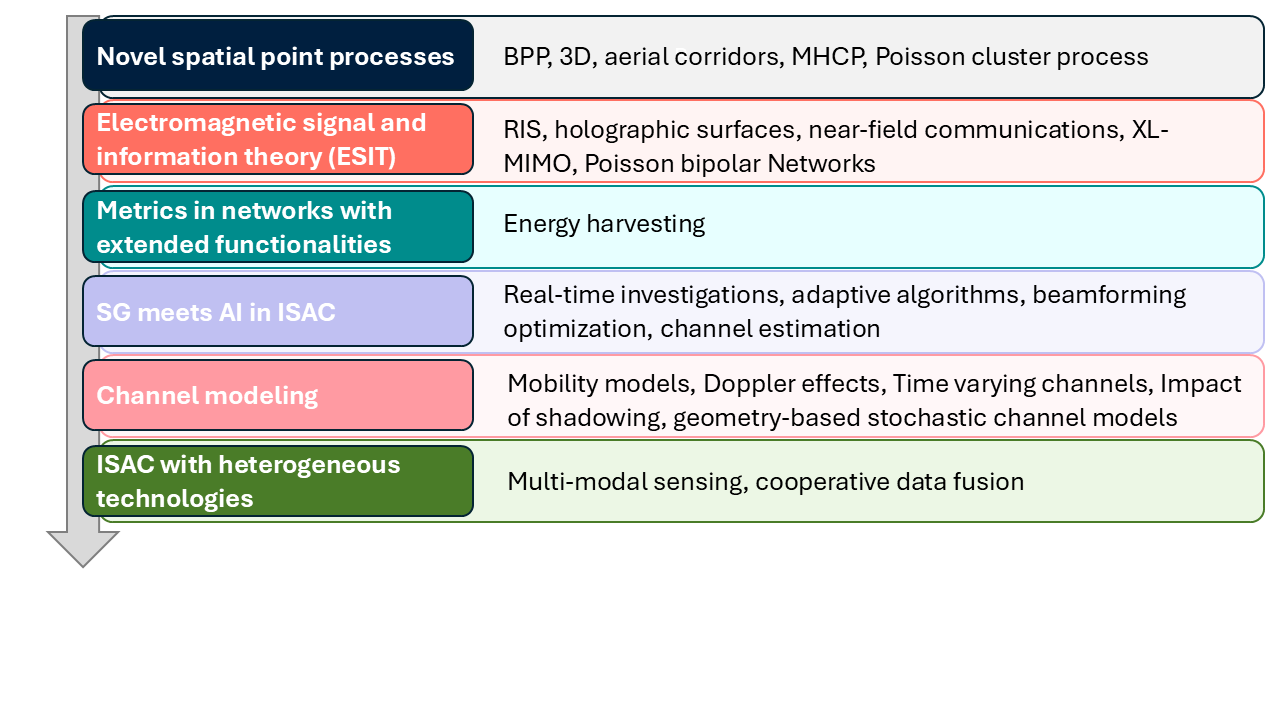}    \vspace{-1.9cm}    \caption{{Future directions}.}

    \label{fig:future}
\end{figure*}

\subsection{Key Results}
According to the area of interest presented in Table \ref{tab:key_results}, we are led to the following concluding remarks:
\begin{itemize}
    \item In the context of SG, meaningful insights have been obtained about the deployment density of RUs, BSs, and blockages and their effect on the performance of communication and sensing networks is demonstrated. 
    \item The trade-off between communication and sensing functionalities in ISAC networks while considering spatial resources and antenna beamwidth design has been examined, and key design guidelines are provided. 
    \item As a key enabling technology for 6G, RISs have been exploited to enhance the JCAS coverage rate of this kind of networks. 
     \item Performance comparison has been carried out between multiple access schemes and between cooperative and standalone sensing.  
     \item  ISAC technology has been used for collision avoidance in UAV networks. 
     \item ISAC-based schemes have been shown to be effective in minimizing interference.
\end{itemize}
Based on the above, it is clear that a notable research interest is directed towards: 1) the effect of deployment density of network entities in the ISAC performance, 2) the effect of directional antenna patterns on the performance of ISAC networks, and 3) the role of enabling technologies and multiple access schemes in mitigating interference and enhancing performance in ISAC networks. Table \ref{tab:key_results} provides a comprehensive summary of the most valuable insights obtained from journal articles\footnote{Each paper has its own focus, merits, and contributions. A paper not listed in Table \ref{tab:key_results} does not mean that the overall contribution of the paper is disparaged in any sense.}.

\subsection{Open Problems and Future Research Directions}

There are still many open research challenges and future directions that require critical attention to drive further progress and meet the required performance goals in communication and sensing enabled networks. In the following, we discuss the identified key open challenges and provide research directions that require further investigation.

\subsubsection{Novel Spatial Point Processes}
The literature review revealed a strong trend in the use of the HPPP to model the spatial locations of network entities. Despite its undoubtedly great potential to maintain analytical tractability, the HPPP has certain limitations. For example, when a predetermined or known number of nodes is deployed in a finite region, the HPPP is definitely not an appropriate choice and the BPP should be utilized instead. Note that the BPP has not yet been used to analyze the performance of ISAC networks. The BPP can reasonably approximate the spatial locations of i) terrestrial UEs in the near-field region of an ISAC-enabled BS (see Fig. 6), ii) UAVs in 2D aerial lanes and iii) UAVs in 3D aerial corridors \cite{10487029,10594734}. Nevertheless, BPP is a non-stationary point process, and therefore it may result in an intractable performance analysis, especially for UAV-enabled ISAC networks. If one is interested in modeling the positions of UAVs in aerial corridors while maintaining a minimum safety distance \cite{9350211}, the MHCP is a reasonable choice (see Fig. 6). For terrestrial networks, the scatterers around a sensing target that cause clutter interference can be reasonably modeled using Poisson cluster processes (see Fig. 6). 

Currently, there is considerable research interest in investigating fundamental performance limits for terrestrial and aerial ISAC networks. However, the research is still in its early stages. The adoption of more realistic spatial point processes for modeling nodes in terrestrial and especially UAV-enabled ISAC networks can provide new insights for fundamental performance limits and key performance metrics. However, the adoption of more realistic spatial point processes may lead to complicated SG-based frameworks. Indeed, the choice of an appropriate spatial point process is not a straightforward task and should be used judiciously to balance the trade-off between complexity and accuracy. The vision of a unified SG-based spatial model for ISAC networks is presented in Fig. 6. 

\begin{figure*}[t!]
  \centering
    \includegraphics[width=7in]{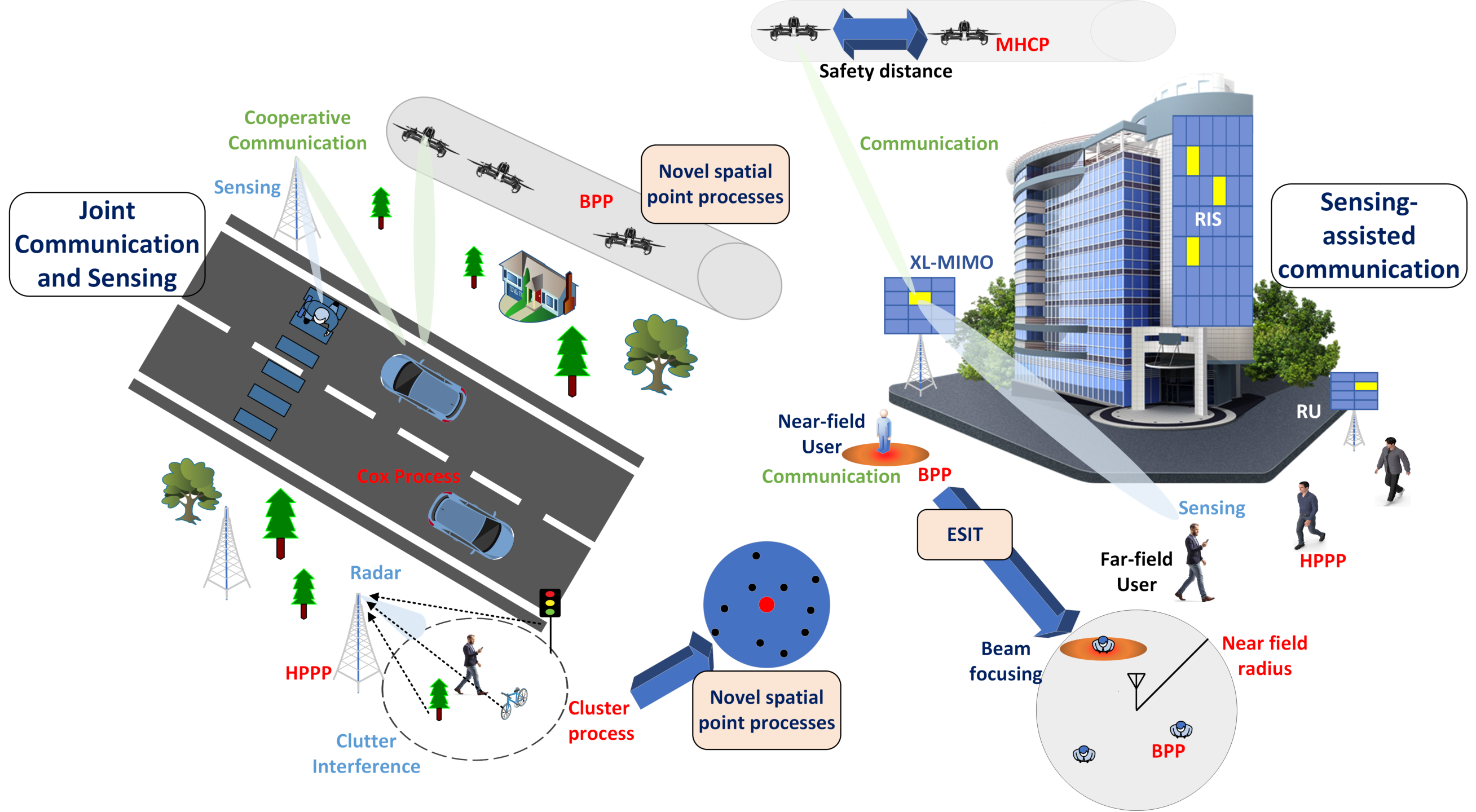}
    \caption{Conceptualization of a unified SG-based framework for communication and sensing networks.}
    \label{fig:SGmodel}
\end{figure*}

\subsubsection{Electromagnetic Signal and Information Theory (ESIT)} ESIT can be defined as physics-aware information theory and signal processing for communications \cite{10415512}. In particular, ESIT is directed to the study of physically consistent schemes for the transmission and processing of information in communication networks with the use of radiating and capable of transmitting and/or receiving information electromagnetic surfaces. 

Very few research studies have considered the promising capabilities of holographic surfaces and HMIMO to ISAC, which we refer to as holographic ISAC (HISAC). These studies demonstrate its potential to improve both sensing and communication performance through optimized beamforming strategies. However, research on HISAC is still in its early stages. In this context, SG can serve as a valuable tool to perform system-level performance analysis of holographic ISAC.

In addition, the large size of emerging antenna technologies, including RIS and holographic surfaces, as well as the potential use of high-frequency bands, makes communications in the near-field region feasible. Although ESIT and near-field communications are undoubtedly an exciting area of research in the context of ISAC technology \cite{10855334,10135096,10498098}, the research gap between SG theory and ESIT remains extremely large. 

In fact, even the role of SG in the performance of near-field communications has barely been revealed \cite{10262267}. The major research challenges preventing the integration of SG into ESIT are related to the physical and electromagnetic properties of the antenna arrays:
\begin{itemize}
    \item The fundamental performance limits in ESIT are strongly based on the characteristics of the antenna arrays, such as the number of elements, the type of aperture array (discrete or continuous), and the orientation of the array. 
     \item The near-field distance-dependent path loss between a UE and an ISAC-enabled BS is strongly dependent on the elements of the antenna arrays. 
     \item Both discrete and continuous apertures are now an integral part of the small-scale fading channel model. In fact, statistical channel models for near-field multipath fading that integrate the properties of continuous-aperture antennas remain an open problem. The main reason for this is that the channel models for continuous aperture antennas are based on the Green function. However, the use of Green's function in scattering environments is extremely challenging \cite{10220205}. 
\end{itemize}
As a result, no system-level performance analysis has been investigated by exploiting the spatial randomness of nodes in ISAC networks while considering the physical properties implied by ESIT. In this direction, new spatial models and channel statistics should be investigated to facilitate the derivation of expressions for key performance metrics. Under the umbrella of SG, possible future directions include the following 
\begin{itemize}
    \item \emph{Poisson bipolar networks:} While Poisson bipolar provides the only spatial model for capturing the orientation of antenna arrays, it may lead to a simple yet reasonable step toward near-field ISAC performance analysis.
    \item\emph{Association policy based on the maximum number of effective degrees of freedom:} Similar to the conventional minimum distance association criterion, novel association policies that integrate ESIT features, e.g. spatial multiplexing gain (degrees of freedom), are expected to provide new insights into fundamental performance limitations. Accordingly, the communicating UE may now associate with the BS providing the maximum number of degrees of freedom.
\end{itemize}
A vision of a hybrid near-field/far-field SG-based spatial model is illustrated in Fig. 6, where the locations of near-field communicating UEs can be modeled as a BPP around an ISAC-enabled BS. The BS can use beam focusing for communication and beamforming for sensing.   

\subsubsection{Metrics in networks with extended functionalities} Considering the demanding key performance indicators of future 6G networks along with the recent advances in ISAC and power transfer capabilities, it is expected that future 6G networks will integrate SWIPT and ISAC. In this direction, new multi-functional networks that can simultaneously provide sensing, communication and power capabilities are currently under remarkable investigation \cite{10382465,10812799}. In the context of SG, recent research focuses on frameworks for the joint integration of communication and energy harvesting functionalities \cite{10382692,10711847}.  Inspired by existing ISAC-related performance metrics, such as the JCAS coverage probability, a new performance metric may arise that goes beyond ISAC, namely the joint coverage probability of communication, sensing, and energy harvesting. Formally, the proposed metric is defined as the probability of achieving the JCAS coverage probability at a given node beyond some target thresholds, given that the energy harvested at the same node exceeds a predefined threshold required for circuit activation. Such kind of extended performance metrics share the vision of future multi-functional 6G networks and pave the way for the formulation of multi-functional SG-based frameworks. Accordingly, analytical expressions for the extended performance bounds will subsequently reveal novel insights and design guidelines for multi-functional 6G networks.   
     
\subsubsection{SG Meets AI in ISAC} As the complexity of ISAC networks continuously increases, AI and ML algorithms are becoming the dominant tools to obtain system-level insights into key ISAC network performance metrics. While SG is limited by a fundamental trade-off between accuracy and complexity, AI is now complementing SG theory to overcome these limitations. ML methods can be used to develop adaptive algorithms that dynamically optimize network performance based on real-time conditions \cite{10578306}. Future studies could explore hybrid strategies that employ predictive beamforming optimization \cite{liu2022learning}, interference management in multi-user ISAC scenarios \cite{10113889}, channel estimation for the sensing and communication signals \cite{10229204}, under the assumption of randomly spatial distributed nodes, targets, scatterers, and/or blockages.

\subsubsection{Channel Modeling} Several aspects of ISAC channel modeling have not been investigated in scenarios where spatial randomness is considered. For example, in most of the presented studies, channel behavior is considered to be slowly varying, even in V2X communications. However, in dynamic networks where mobile user tracking is performed, e.g., \cite{mao2024uav}, SG tools can increase the reliability of performance results. In addition, in time-varying scenarios, the Doppler effect should not be ignored, as it can severely affect channel behavior \cite{favarelli2025sensor}. Furthermore, the impact of random shadowing in ISAC scenarios has not been thoroughly investigated, e.g., \cite{bithas2024joint}, especially by using SG tools. Finally, SG tools could be very helpful for presenting novel geometry-based stochastic channel models in ISAC scenarios \cite{pradhan2023stochastic}.

\subsubsection{ISAC with Heterogeneous Technologies} Multi-sensor cooperative perception with diverse sensor types is a key challenge. Although many ISAC systems focus on a single sensor type, such as radar, future large-scale networks will utilize more diverse sensor technologies, such as cameras, light-emitting diodes (LEDs), lidars, or thermal imagers \cite{10716723,bozanis2024}. SG methods can play a critical role in capturing how these different sensors, each with its own field of view, detection range, and clutter sensitivity, are distributed over large areas. For example, cameras can provide rich visual detail but only at certain viewing angles, while radar can penetrate certain obstacles but may be susceptible to specular reflections. By building a unified SG-based model that collectively represents the deployment and coverage areas of these sensors, the coverage probability, detection reliability, and accuracy can be quantified. Key challenges include i) the determination of the best approach to cluster or schedule sensors to minimize blind spots and avoid redundant scanning, ii) the establishment of cooperative data fusion rules that account for varying accuracy and update rates, and iii) mitigation of heterogeneous interference or blocking effects. Investigating these issues with SG will provide critical design insights, which range from the identification of the different sensors deployment for optimal joint perception to manage the communication overhead of fusing sensor data at the network edge.

\section{Conclusions}
SG has long been a powerful tool for evaluating the performance of complex wireless networks. By modeling spatial locations through point processes, a reliable analytical performance evaluation can be performed and used to provide useful design insights. Despite the obvious usefulness of this tool in scenarios where the location of network elements is very important for performance evaluation, it has only recently been adopted in ISAC network studies. In this survey, a comprehensive analysis of the technical literature on SG for sensing and communication modeling and analysis was conducted.
and communication. The article started with a short tutorial on the SG tools and the basic concepts of communication- and sensing-enabled networks. In this context, a subsection with performance metrics used in the literature was presented. Then, system models and key technical components used in the literature for terrestrial, vehicular, and aerial networks were reviewed, highlighting the main results achieved so far in JCAS, SAC, and CAS networks. Finally, the article identified open problems and future research directions in SG for ISAC, many of which remain unexplored.
\balance
\bibliographystyle{IEEEtran}
\bibliography{IEEEabrv,references}

\begin{thebibliography}{100}
\providecommand{\url}[1]{#1}
\csname url@samestyle\endcsname
\providecommand{\newblock}{\relax}
\providecommand{\bibinfo}[2]{#2}
\providecommand{\BIBentrySTDinterwordspacing}{\spaceskip=0pt\relax}
\providecommand{\BIBentryALTinterwordstretchfactor}{4}
\providecommand{\BIBentryALTinterwordspacing}{\spaceskip=\fontdimen2\font plus
\BIBentryALTinterwordstretchfactor\fontdimen3\font minus \fontdimen4\font\relax}
\providecommand{\BIBforeignlanguage}[2]{{%
\expandafter\ifx\csname l@#1\endcsname\relax
\typeout{** WARNING: IEEEtran.bst: No hyphenation pattern has been}%
\typeout{** loaded for the language `#1'. Using the pattern for}%
\typeout{** the default language instead.}%
\else
\language=\csname l@#1\endcsname
\fi
#2}}
\providecommand{\BIBdecl}{\relax}
\BIBdecl

\bibitem{9040264}
M.~Giordani, M.~Polese, M.~Mezzavilla, S.~Rangan, and M.~Zorzi, ``Toward {6G} networks: Use cases and technologies,'' \emph{IEEE Communications Magazine}, vol.~58, no.~3, pp. 55--61, 2020.

\bibitem{recommendation2023framework}
I.~Recommendation, ``Framework and overall objectives of the future development of imt for 2030 and beyond,'' \emph{International Telecommunication Union (ITU) Recommendation (ITU-R)}, 2023.

\bibitem{jiang2021road}
W.~Jiang, B.~Han, M.~A. Habibi, and H.~D. Schotten, ``The road towards {6G}: A comprehensive survey,'' \emph{IEEE Open Journal of the Communications Society}, vol.~2, pp. 334--366, 2021.

\bibitem{wymeersch2021integration}
H.~Wymeersch, D.~Shrestha, C.~M. De~Lima, V.~Yajnanarayana, B.~Richerzhagen, M.~F. Keskin, K.~Schindhelm, A.~Ramirez, A.~Wolfgang, M.~F. De~Guzman \emph{et~al.}, ``Integration of communication and sensing in {6G}: A joint industrial and academic perspective,'' in \emph{2021 IEEE 32nd Annual International Symposium on Personal, Indoor and Mobile Radio Communications (PIMRC)}.\hskip 1em plus 0.5em minus 0.4em\relax IEEE, 2021, pp. 1--7.

\bibitem{tan2021integrated}
D.~K.~P. Tan, J.~He, Y.~Li, A.~Bayesteh, Y.~Chen, P.~Zhu, and W.~Tong, ``Integrated sensing and communication in {6G}: Motivations, use cases, requirements, challenges and future directions,'' in \emph{2021 1st IEEE International Online Symposium on Joint Communications \& Sensing (JC\&S)}.\hskip 1em plus 0.5em minus 0.4em\relax IEEE, 2021, pp. 1--6.

\bibitem{special}
\BIBentryALTinterwordspacing
{IEEE Journal on Selected Areas in Communications}. (2025) Recent advances in integrated sensing and communications. [Online]. Available: \url{https://www.comsoc.org/publications/journals/ieee-jsac/cfp/recent-advances-integrated-sensing-and-communications}
\BIBentrySTDinterwordspacing

\bibitem{ISAC_ITE}
\BIBentryALTinterwordspacing
{IEEE Communications Sociaty}. (2025) Integrated sensing and communication ({ISAC}) emerging technology initiative. [Online]. Available: \url{https://isac.committees.comsoc.org/}
\BIBentrySTDinterwordspacing

\bibitem{baccelli1997stochastic}
F.~Baccelli, M.~Klein, M.~Lebourges, and S.~Zuyev, ``Stochastic geometry and architecture of communication networks,'' \emph{Telecommunication Systems}, vol.~7, pp. 209--227, 1997.

\bibitem{baccelli1998stochastic}
F.~Baccelli and S.~Zuyev, ``Stochastic geometry models of mobile communication networks,'' in \emph{Frontiers in queueing: models and applications in science and engineering}, 1998, pp. 227--243.

\bibitem{8742579}
A.~Zappone, M.~Di~Renzo, and M.~Debbah, ``Wireless networks design in the era of deep learning: Model-based, ai-based, or both?'' \emph{IEEE Transactions on Communications}, vol.~67, no.~10, pp. 7331--7376, 2019.

\bibitem{8765703}
A.~Zappone, M.~Di~Renzo, M.~Debbah, T.~T. Lam, and X.~Qian, ``Model-aided wireless artificial intelligence: Embedding expert knowledge in deep neural networks for wireless system optimization,'' \emph{IEEE Vehicular Technology Magazine}, vol.~14, no.~3, pp. 60--69, 2019.

\bibitem{8885526}
B.~Błaszczyszyn and H.~Keeler, ``Determinantal thinning of point processes with network learning applications,'' in \emph{2019 IEEE Wireless Communications and Networking Conference (WCNC)}, 2019, pp. 1--8.

\bibitem{saha2019machine}
C.~Saha and H.~S. Dhillon, ``Machine learning meets stochastic geometry: Determinantal subset selection for wireless networks,'' in \emph{2019 IEEE Global Communications Conference (GLOBECOM)}.\hskip 1em plus 0.5em minus 0.4em\relax IEEE, 2019, pp. 1--6.

\bibitem{10418473}
S.~Lu, F.~Liu, Y.~Li, K.~Zhang, H.~Huang, J.~Zou, X.~Li, Y.~Dong, F.~Dong, J.~Zhu, Y.~Xiong, W.~Yuan, Y.~Cui, and L.~Hanzo, ``Integrated sensing and communications: Recent advances and ten open challenges,'' \emph{IEEE Internet of Things Journal}, vol.~11, no.~11, pp. 19\,094--19\,120, 2024.

\bibitem{9393464}
N.~C. Luong, X.~Lu, D.~T. Hoang, D.~Niyato, and D.~I. Kim, ``Radio resource management in joint radar and communication: A comprehensive survey,'' \emph{IEEE Communications Surveys \& Tutorials}, vol.~23, no.~2, pp. 780--814, 2021.

\bibitem{10812728}
D.~Wen, Y.~Zhou, X.~Li, Y.~Shi, K.~Huang, and K.~B. Letaief, ``A survey on integrated sensing, communication, and computation,'' \emph{IEEE Communications Surveys \& Tutorials}, pp. 1--1, 2024.

\bibitem{8667902}
A.~Sarker, H.~Shen, M.~Rahman, M.~Chowdhury, K.~Dey, F.~Li, Y.~Wang, and H.~S. Narman, ``A review of sensing and communication, human factors, and controller aspects for information-aware connected and automated vehicles,'' \emph{IEEE Transactions on Intelligent Transportation Systems}, vol.~21, no.~1, pp. 7--29, 2020.

\bibitem{9705498}
A.~Liu, Z.~Huang, M.~Li, Y.~Wan, W.~Li, T.~X. Han, C.~Liu, R.~Du, D.~K.~P. Tan, J.~Lu, Y.~Shen, F.~Colone, and K.~Chetty, ``A survey on fundamental limits of integrated sensing and communication,'' \emph{IEEE Communications Surveys \& Tutorials}, vol.~24, no.~2, pp. 994--1034, 2022.

\bibitem{10756650}
M.~Ade Krisna~Respati and B.~M. Lee, ``A survey on machine learning enhanced integrated sensing and communication systems: Architectures, algorithms, and applications,'' \emph{IEEE Access}, vol.~12, pp. 170\,946--170\,964, 2024.

\bibitem{9330512}
C.~De~Lima, D.~Belot, R.~Berkvens, A.~Bourdoux, D.~Dardari, M.~Guillaud, M.~Isomursu, E.-S. Lohan, Y.~Miao, A.~N. Barreto, M.~R.~K. Aziz, J.~Saloranta, T.~Sanguanpuak, H.~Sarieddeen, G.~Seco-Granados, J.~Suutala, T.~Svensson, M.~Valkama, B.~Van~Liempd, and H.~Wymeersch, ``Convergent communication, sensing and localization in {6G} systems: An overview of technologies, opportunities and challenges,'' \emph{IEEE Access}, vol.~9, pp. 26\,902--26\,925, 2021.

\bibitem{9585321}
J.~A. Zhang, M.~L. Rahman, K.~Wu, X.~Huang, Y.~J. Guo, S.~Chen, and J.~Yuan, ``Enabling joint communication and radar sensing in mobile networks—a survey,'' \emph{IEEE Communications Surveys \& Tutorials}, vol.~24, no.~1, pp. 306--345, 2022.

\bibitem{10646523}
Z.~Wei, J.~Jia, Y.~Niu, L.~Wang, H.~Wu, H.~Yang, and Z.~Feng, ``Integrated sensing and communication channel modeling: A survey,'' \emph{IEEE Internet of Things Journal}, pp. 1--1, 2024.

\bibitem{9829746}
J.~Wang, N.~Varshney, C.~Gentile, S.~Blandino, J.~Chuang, and N.~Golmie, ``Integrated sensing and communication: Enabling techniques, applications, tools and data sets, standardization, and future directions,'' \emph{IEEE Internet of Things Journal}, vol.~9, no.~23, pp. 23\,416--23\,440, 2022.

\bibitem{10012421}
Z.~Wei, H.~Qu, Y.~Wang, X.~Yuan, H.~Wu, Y.~Du, K.~Han, N.~Zhang, and Z.~Feng, ``Integrated sensing and communication signals toward {5G-A} and {6G}: A survey,'' \emph{IEEE Internet of Things Journal}, vol.~10, no.~13, pp. 11\,068--11\,092, 2023.

\bibitem{9924202}
W.~Zhou, R.~Zhang, G.~Chen, and W.~Wu, ``Integrated sensing and communication waveform design: A survey,'' \emph{IEEE Open Journal of the Communications Society}, vol.~3, pp. 1930--1949, 2022.

\bibitem{10608156}
X.~Zhu, J.~Liu, L.~Lu, T.~Zhang, T.~Qiu, C.~Wang, and Y.~Liu, ``Enabling intelligent connectivity: A survey of secure {ISAC} in {6G} networks,'' \emph{IEEE Communications Surveys \& Tutorials}, pp. 1--1, 2024.

\bibitem{5226957}
M.~Haenggi, J.~G. Andrews, F.~Baccelli, O.~Dousse, and M.~Franceschetti, ``Stochastic geometry and random graphs for the analysis and design of wireless networks,'' \emph{IEEE Journal on Selected Areas in Communications}, vol.~27, no.~7, pp. 1029--1046, 2009.

\bibitem{7733098}
H.~ElSawy, A.~Sultan-Salem, M.-S. Alouini, and M.~Z. Win, ``Modeling and analysis of cellular networks using stochastic geometry: A tutorial,'' \emph{IEEE Communications Surveys \& Tutorials}, vol.~19, no.~1, pp. 167--203, 2017.

\bibitem{6524460}
H.~ElSawy, E.~Hossain, and M.~Haenggi, ``Stochastic geometry for modeling, analysis, and design of multi-tier and cognitive cellular wireless networks: A survey,'' \emph{IEEE Communications Surveys \& Tutorials}, vol.~15, no.~3, pp. 996--1019, 2013.

\bibitem{9516701}
X.~Lu, M.~Salehi, M.~Haenggi, E.~Hossain, and H.~Jiang, ``Stochastic geometry analysis of spatial-temporal performance in wireless networks: A tutorial,'' \emph{IEEE Communications Surveys \& Tutorials}, vol.~23, no.~4, pp. 2753--2801, 2021.

\bibitem{9378781}
Y.~Hmamouche, M.~Benjillali, S.~Saoudi, H.~Yanikomeroglu, and M.~D. Renzo, ``New trends in stochastic geometry for wireless networks: A tutorial and survey,'' \emph{Proceedings of the IEEE}, vol. 109, no.~7, pp. 1200--1252, 2021.

\bibitem{10574257}
R.~Wang, B.~E.~Y. Belmekki, X.~Zhang, and M.-S. Alouini, ``Network-level analysis of integrated sensing and communication using stochastic geometry,'' \emph{IEEE Internet of Things Magazine}, vol.~7, no.~4, pp. 84--90, 2024.

\bibitem{haenggi2013stochastic}
M.~Haenggi, \emph{Stochastic geometry for wireless networks}.\hskip 1em plus 0.5em minus 0.4em\relax Cambridge University Press, 2013.

\bibitem{10556618}
X.~Gan, C.~Huang, Z.~Yang, X.~Chen, J.~He, Z.~Zhang, C.~Yuen, Y.~Liang~Guan, and M.~Debbah, ``Coverage and rate analysis for integrated sensing and communication networks,'' \emph{IEEE Journal on Selected Areas in Communications}, vol.~42, no.~9, pp. 2213--2227, 2024.

\bibitem{9538929}
X.~Liang, Y.~Zhong, and L.~Zou, ``Real-time sensing and communication improvement with mobile sensors,'' in \emph{2021 IEEE/CIC International Conference on Communications in China (ICCC Workshops)}, 2021, pp. 71--76.

\bibitem{10565853}
A.~S. Parihar, K.~Singh, V.~Bhatia, C.-P. Li, and T.~Q. Duong, ``Performance analysis of {NOMA}-enabled active {RIS}-aided {MIMO} heterogeneous iot networks with integrated sensing and communication,'' \emph{IEEE Internet of Things Journal}, vol.~11, no.~17, pp. 28\,137--28\,152, 2024.

\bibitem{10633859}
W.~Chen, L.~Li, Z.~Chen, Y.~Liu, B.~Ning, and T.~Q.~S. Quek, ``{ISAC}-enabled beam alignment for terahertz networks: Scheme design and coverage analysis,'' \emph{IEEE Transactions on Vehicular Technology}, vol.~73, no.~12, pp. 19\,019--19\,033, 2024.

\bibitem{9893396}
I.~Tropkina, B.~Sun, D.~Moltchanov, A.~Pyattaev, B.~Tan, R.~Dinis, and S.~Andreev, ``Distributed communication and sensing system co-design for improved {UAV} network resilience,'' \emph{IEEE Transactions on Vehicular Technology}, vol.~72, no.~1, pp. 924--939, 2023.

\bibitem{10412651}
M.~Mei, M.~Yao, Q.~Yang, J.~Wang, and R.~R. Rao, ``Stochastic network calculus analysis of spatial-temporal integrated sensing and communication networks,'' \emph{IEEE Transactions on Vehicular Technology}, vol.~73, no.~6, pp. 9120--9124, 2024.

\bibitem{10681925}
S.~Hu, J.~Gao, X.~Huang, C.~Zhou, M.~He, and X.~S. Shen, ``Model drift-adaptive resource reservation in {ISAC} networks: A digital twin-based approach,'' in \emph{2024 IEEE/CIC International Conference on Communications in China (ICCC)}, 2024, pp. 2143--2148.

\bibitem{7155510}
Y.~Li, F.~Baccelli, H.~S. Dhillon, and J.~G. Andrews, ``Statistical modeling and probabilistic analysis of cellular networks with determinantal point processes,'' \emph{IEEE Transactions on Communications}, vol.~63, no.~9, pp. 3405--3422, 2015.

\bibitem{lavancier2012statistical}
F.~Lavancier, J.~M{\o}ller, and E.~Rubak, ``Statistical aspects of determinantal point processes,'' 2012.

\bibitem{10603243}
T.~R, M.~T, K.~Ka, R.~J, and A.~G, ``Success probability analysis for joint sensing and communications: A determinantal point process perspective,'' in \emph{2024 International Conference on Integrated Circuits, Communication, and Computing Systems (ICIC3S)}, vol.~1, 2024, pp. 1--6.

\bibitem{9420372}
C.~Skouroumounis, C.~Psomas, and I.~Krikidis, ``Fd-jcas techniques for mmwave hetnets: Ginibre point process modeling and analysis,'' \emph{IEEE Transactions on Mobile Computing}, vol.~21, no.~12, pp. 4352--4366, 2022.

\bibitem{9814629}
N.~Jiang, S.~Yan, Z.~Liu, C.~Hu, and M.~Peng, ``Communication and computation assisted sensing information freshness performance analysis in vehicular networks,'' in \emph{2022 IEEE International Conference on Communications Workshops (ICC Workshops)}, 2022, pp. 969--974.

\bibitem{10582835}
Y.~Wang, Q.~Zhang, J.~Andrew~Zhang, Z.~Wei, Z.~Feng, and J.~Peng, ``Interference characterization and mitigation for multi-beam {ISAC} systems in vehicular networks,'' \emph{IEEE Transactions on Wireless Communications}, vol.~23, no.~10, pp. 14\,729--14\,742, 2024.

\bibitem{8828016}
L.~Zheng, M.~Lops, Y.~C. Eldar, and X.~Wang, ``Radar and communication coexistence: An overview: A review of recent methods,'' \emph{IEEE Signal Processing Magazine}, vol.~36, no.~5, pp. 85--99, 2019.

\bibitem{wu2022joint}
K.~Wu, J.~A. Zhang, and Y.~J. Guo, \emph{Joint communications and sensing: From fundamentals to advanced techniques}.\hskip 1em plus 0.5em minus 0.4em\relax John Wiley \& Sons, 2022.

\bibitem{10845869}
F.~Dong, F.~Liu, S.~Lu, Y.~Xiong, Q.~Zhang, Z.~Feng, and F.~Gao, ``Communication-assisted sensing in {6G} networks,'' \emph{IEEE Journal on Selected Areas in Communications}, pp. 1--1, 2025.

\bibitem{10422712}
Q.~Xue, C.~Ji, S.~Ma, J.~Guo, Y.~Xu, Q.~Chen, and W.~Zhang, ``A survey of beam management for mmwave and {THz} communications towards {6G},'' \emph{IEEE Communications Surveys \& Tutorials}, vol.~26, no.~3, pp. 1520--1559, 2024.

\bibitem{10562219}
J.~Xu, M.~A. Kishk, J.~P. Coon, and M.-S. Alouini, ``Performance analysis and optimal resource allocation for large scale joint sensing and communication,'' \emph{IEEE Transactions on Wireless Communications}, vol.~23, no.~10, pp. 14\,350--14\,364, 2024.

\bibitem{10569084}
F.~D.~S. Moulin, C.~Wiame, L.~Vandendorpe, and C.~Oestges, ``Joint coverage and detection performance metrics for integrated sensing and communication in automotive scenarios,'' \emph{IEEE Transactions on Vehicular Technology}, vol.~73, no.~11, pp. 16\,758--16\,773, 2024.

\bibitem{10568512}
S.~Tripathi, A.~K. Gupta, and S.~Amuru, ``On the coverage of cognitive mmwave networks with directional sensing and communication,'' \emph{IEEE Transactions on Wireless Communications}, vol.~23, no.~10, pp. 14\,215--14\,231, 2024.

\bibitem{10433485}
Y.~Nabil, H.~Elsawy, S.~Al-Dharrab, H.~Mostafa, and H.~Attia, ``Beamwidth design tradeoffs in radar-aided millimeter-wave cellular networks: A stochastic geometry approach,'' \emph{IEEE Access}, vol.~12, pp. 26\,196--26\,211, 2024.

\bibitem{10320397}
N.~R. Olson, J.~G. Andrews, and R.~W. Heath, ``Coverage and rate of joint communication and parameter estimation in wireless networks,'' \emph{IEEE Transactions on Information Theory}, vol.~70, no.~1, pp. 206--243, 2024.

\bibitem{10735119}
K.~Meng, C.~Masouros, G.~Chen, and F.~Liu, ``Network-level integrated sensing and communication: Interference management and {BS} coordination using stochastic geometry,'' \emph{IEEE Transactions on Wireless Communications}, vol.~23, no.~12, pp. 19\,365--19\,381, 2024.

\bibitem{10490156}
A.~Salem, K.~Meng, C.~Masouros, F.~Liu, and D.~Lopez-Perez, ``Rethinking dense cells for integrated sensing and communications: A stochastic geometric view,'' \emph{IEEE Open Journal of the Communications Society}, vol.~5, pp. 2226--2239, 2024.

\bibitem{10615428}
K.~Ma, C.~Feng, G.~Geraci, and H.~H. Yang, ``The meta distribution of the sir in joint communication and sensing networks,'' in \emph{2024 IEEE International Conference on Communications Workshops (ICC Workshops)}, 2024, pp. 691--696.

\bibitem{10747175}
Y.~Zhang, H.~Shan, H.~Chen, L.~Cai, Z.~Shi, T.~Q.~S. Quek, and L.~Sheng, ``Perceptive mobile networks for standalone and cooperative {UAV} surveillance,'' \emph{IEEE Transactions on Wireless Communications}, vol.~23, no.~12, pp. 19\,916--19\,932, 2024.

\bibitem{ram2022optimization}
S.~S. Ram, S.~Singhal, and G.~Ghatak, ``Optimization of network throughput of joint radar communication system using stochastic geometry,'' \emph{Frontiers in Signal Processing}, vol.~2, p. 835743, 2022.

\bibitem{10570658}
K.~S. Ali, R.~Bomfin, and M.~Chafii, ``Successive interference cancellation for {ISAC} in a large full-duplex cellular network,'' in \emph{2024 IEEE Wireless Communications and Networking Conference (WCNC)}, 2024, pp. 1--6.

\bibitem{9931437}
H.~Ma, Z.~Wei, Z.~Li, F.~Ning, X.~Chen, and Z.~Feng, ``Performance of cooperative detection in joint communication-sensing vehicular network: A data analytic and stochastic geometry approach,'' \emph{IEEE Transactions on Vehicular Technology}, vol.~72, no.~3, pp. 3848--3863, 2023.

\bibitem{10217344}
S.~J. Maeng, J.~Park, and I.~Guvenc, ``Analysis of {UAV} radar and communication network coexistence with different multiple access protocols,'' \emph{IEEE Transactions on Communications}, vol.~71, no.~11, pp. 6578--6592, 2023.

\bibitem{10695883}
X.~Gan, C.~Huang, Z.~Yang, X.~Chen, F.~Bader, Z.~Zhang, C.~Yuen, Y.~L. Guan, and M.~Debbah, ``{RIS}-assisted coverage enhancement in mmwave integrated sensing and communication networks,'' in \emph{2024 International Conference on Ubiquitous Communication (Ucom)}, 2024, pp. 21--26.

\bibitem{dahlman20205g}
E.~Dahlman, S.~Parkvall, and J.~Skold, \emph{{5G} NR: The next generation wireless access technology}.\hskip 1em plus 0.5em minus 0.4em\relax Academic Press, 2020.

\bibitem{9765510}
W.~Chen, L.~Li, Z.~Chen, T.~Quek, and S.~Li, ``Enhancing {THz}/mmwave network beam alignment with integrated sensing and communication,'' \emph{IEEE Communications Letters}, vol.~26, no.~7, pp. 1698--1702, 2022.

\bibitem{10694260}
K.~Meng and C.~Masouros, ``Cooperative sensing and communication for {ISAC} networks: Performance analysis and optimization,'' in \emph{2024 IEEE 25th International Workshop on Signal Processing Advances in Wireless Communications (SPAWC)}, 2024, pp. 446--450.

\bibitem{10769538}
K.~Meng, C.~Masouros, A.~P. Petropulu, and L.~Hanzo, ``Cooperative {ISAC} networks: Performance analysis, scaling laws and optimization,'' \emph{IEEE Transactions on Wireless Communications}, pp. 1--1, 2024.

\bibitem{10278626}
Y.~Nabil, H.~ElSawy, S.~Al–Dharrab, H.~Attia, and H.~Mostafa, ``A stochastic geometry analysis for joint radar communication system in millimeter-wave band,'' in \emph{ICC 2023 - IEEE International Conference on Communications}, 2023, pp. 5849--5854.

\bibitem{10683162}
X.~Gan, C.~Huang, Z.~Yang, X.~Chen, J.~He, Z.~Zhang, C.~Yuen, Y.~L. Guan, and M.~Debbah, ``Toward a unified analytical framework for {ISAC} fundamentals in cellular networks,'' in \emph{2024 IEEE 99th Vehicular Technology Conference (VTC2024-Spring)}, 2024, pp. 1--6.

\bibitem{10622535}
W.~Cheng, Z.~Zhao, H.~H. Yang, W.~Hong, T.~Q. Quek, and Z.~Ding, ``On the study of success serving probability for integrated sensing and communication ({ISAC}) based on stochastic geometry,'' in \emph{ICC 2024 - IEEE International Conference on Communications}, 2024, pp. 5098--5103.

\bibitem{10579074}
Z.~Sun, S.~Yan, N.~Jiang, J.~Zhou, and M.~Peng, ``Performance analysis of integrated sensing and communication networks with blockage effects,'' \emph{IEEE Transactions on Vehicular Technology}, vol.~73, no.~11, pp. 16\,876--16\,891, 2024.

\bibitem{10437749}
W.~Chen, L.~Li, B.~Ning, Z.~Chen, and T.~Q. Quek, ``Sensing resource allocation for enlarging the coverage range of {ISAC}-based terahertz network,'' in \emph{GLOBECOM 2023 - 2023 IEEE Global Communications Conference}, 2023, pp. 3639--3644.

\bibitem{8681732}
S.~Kim, H.~Cha, J.~Kim, S.-W. Ko, and S.-L. Kim, ``Sense-and-predict: Harnessing spatial interference correlation for cognitive radio networks,'' \emph{IEEE Transactions on Wireless Communications}, vol.~18, no.~5, pp. 2777--2793, 2019.

\bibitem{10622748}
Y.~Wu and C.~Han, ``Coverage and capacity analysis for terahertz integrated sensing and communication networks,'' in \emph{ICC 2024 - IEEE International Conference on Communications}, 2024, pp. 3555--3560.

\bibitem{10571293}
N.~Keshtiarast, P.~Kumar~Bishoyi, and M.~Petrova, ``Modeling and performance analysis of csma-based jcas networks,'' in \emph{2024 IEEE Wireless Communications and Networking Conference (WCNC)}, 2024, pp. 01--06.

\bibitem{10436875}
K.~S. Ali and M.~Chafii, ``A dynamic transmission strategy for {ISAC} in large networks,'' in \emph{GLOBECOM 2023 - 2023 IEEE Global Communications Conference}, 2023, pp. 4576--4582.

\bibitem{de2024full}
F.~De~Saint~Moulin, S.~Demey, C.~Wiame, L.~Vandendorpe, and C.~Oestges, ``Full-duplex v2x integrated sensing and communication scenario: Stochastic geometry, monte-carlo, and ray-tracing comparison,'' \emph{arXiv e-prints}, pp. arXiv--2407, 2024.

\bibitem{10682039}
L.~Wang, Y.~Zhang, H.~Shan, C.~Chen, F.~Hou, H.~Ghafoor, and Y.~Cheng, ``Performance analysis and optimization of {ISAC} vehicular networks with 360° radar detection,'' in \emph{2024 IEEE/CIC International Conference on Communications in China (ICCC)}, 2024, pp. 580--585.

\bibitem{10695929}
Y.~Jiang, F.~Meng, X.~Li, X.~Li, G.~Zhu, K.~Han, and Q.~Shi, ``Coverage analysis for air-ground integrated-sensing-and-communication networks,'' in \emph{2024 International Conference on Ubiquitous Communication (Ucom)}, 2024, pp. 455--459.

\bibitem{10376044}
Z.~Liu, F.~Zesong, P.~Liu, X.~Wang, Z.~Zheng, D.~Zhou, and W.~Yuan, ``Integrated sensing and communication for {UAV}-borne {SAR} systems,'' in \emph{2023 22nd International Symposium on Communications and Information Technologies (ISCIT)}, 2023, pp. 1--6.

\bibitem{9847217}
S.~Hu, X.~Yuan, W.~Ni, and X.~Wang, ``Trajectory planning of cellular-connected {UAV} for communication-assisted radar sensing,'' \emph{IEEE Transactions on Communications}, vol.~70, no.~9, pp. 6385--6396, 2022.

\bibitem{9606367}
D.~Ghozlani, A.~Omri, S.~Bouallegue, H.~Chamkhia, and R.~Bouallegue, ``Stochastic geometry-based analysis of joint radar and communication-enabled cooperative detection systems,'' in \emph{2021 17th International Conference on Wireless and Mobile Computing, Networking and Communications (WiMob)}, 2021, pp. 325--330.

\bibitem{huang2019v2x}
J.~Huang, Z.~Fei, T.~Wang, X.~Wang, F.~Liu, H.~Zhou, J.~A. Zhang, and G.~Wei, ``V2x-communication assisted interference minimization for automotive radars,'' \emph{China Communications}, vol.~16, no.~10, pp. 100--111, 2019.

\bibitem{liu2024impacts}
C.~Liu, C.~Liu, W.~Chen, and M.~Peng, ``Impacts of blockages on {UAV}-enabled integrated sensing and communications,'' in \emph{2024 16th International Conference on Wireless Communications and Signal Processing (WCSP)}.\hskip 1em plus 0.5em minus 0.4em\relax IEEE, 2024, pp. 825--831.

\bibitem{meng2024network}
K.~Meng, K.~Han, C.~Masouros, and L.~Hanzo, ``Network-level {ISAC}: Performance analysis and optimal antenna-to-{BS} allocation,'' \emph{arXiv preprint arXiv:2410.06365}, 2024.

\bibitem{10217179}
X.~Li, S.~Guo, T.~Li, X.~Zou, and D.~Li, ``On the performance trade-off of distributed integrated sensing and communication networks,'' \emph{IEEE Wireless Communications Letters}, vol.~12, no.~12, pp. 2033--2037, 2023.

\bibitem{10636567}
K.~Meng and C.~Masouros, ``Networked {ISAC} coordinated beamforming and cooperative {BS} cluster optimization,'' in \emph{2024 14th International Symposium on Communication Systems, Networks and Digital Signal Processing (CSNDSP)}, 2024, pp. 415--420.

\bibitem{8855935}
Z.~Fang, Z.~Wei, Z.~Feng, X.~Chen, and Z.~Guo, ``Performance of joint radar and communication enabled cooperative detection,'' in \emph{2019 IEEE/CIC International Conference on Communications in China (ICCC)}, 2019, pp. 753--758.

\bibitem{9049687}
P.~Ren, A.~Munari, and M.~Petrova, ``Performance analysis of a time-sharing joint radar-communications network,'' in \emph{2020 International Conference on Computing, Networking and Communications (ICNC)}, 2020, pp. 908--913.

\bibitem{10549451}
A.~Sneh and S.~S. Ram, ``Radar operating metrics and network throughput for integrated sensing and communications in millimeter-wave urban environments,'' in \emph{2024 IEEE Radar Conference (RadarConf24)}, 2024, pp. 1--6.

\bibitem{9078878}
M.~Banagar and H.~S. Dhillon, ``Performance characterization of canonical mobility models in drone cellular networks,'' \emph{IEEE Transactions on Wireless Communications}, vol.~19, no.~7, pp. 4994--5009, 2020.

\bibitem{10487029}
S.~J. Maeng and I.~G\"uven\c~c, ``Uav corridor coverage analysis with base station antenna uptilt and strongest signal association,'' \emph{IEEE Transactions on Aerospace and Electronic Systems}, vol.~60, no.~4, pp. 5621--5630, 2024.

\bibitem{10594734}
S.~Karimi-Bidhendi, G.~Geraci, and H.~Jafarkhani, ``Optimizing cellular networks for uav corridors via quantization theory,'' \emph{IEEE Transactions on Wireless Communications}, vol.~23, no.~10, pp. 14\,924--14\,939, 2024.

\bibitem{9350211}
J.~Lyu and H.-M. Wang, ``Secure uav random networks with minimum safety distance,'' \emph{IEEE Transactions on Vehicular Technology}, vol.~70, no.~3, pp. 2856--2861, 2021.

\bibitem{10415512}
M.~D. Renzo and M.~D. Migliore, ``Electromagnetic signal and information theory,'' \emph{IEEE BITS the Information Theory Magazine}, pp. 1--13, 2024.

\bibitem{10855334}
Y.~Jiang, F.~Gao, S.~Jin, and T.~J. Cui, ``Electromagnetic property sensing in isac with multiple base stations: Algorithm, pilot design, and performance analysis,'' \emph{IEEE Transactions on Wireless Communications}, pp. 1--1, 2025.

\bibitem{10135096}
Z.~Wang, X.~Mu, and Y.~Liu, ``Near-field integrated sensing and communications,'' \emph{IEEE Communications Letters}, vol.~27, no.~8, pp. 2048--2052, 2023.

\bibitem{10498098}
B.~Zhao, C.~Ouyang, Y.~Liu, X.~Zhang, and H.~V. Poor, ``Modeling and analysis of near-field isac,'' \emph{IEEE Journal of Selected Topics in Signal Processing}, vol.~18, no.~4, pp. 678--693, 2024.

\bibitem{10262267}
Z.~Xie, Y.~Liu, J.~Xu, X.~Wu, and A.~Nallanathan, ``Performance analysis for near-field mimo: Discrete and continuous aperture antennas,'' \emph{IEEE Wireless Communications Letters}, vol.~12, no.~12, pp. 2258--2262, 2023.

\bibitem{10220205}
Y.~Liu, Z.~Wang, J.~Xu, C.~Ouyang, X.~Mu, and R.~Schober, ``Near-field communications: A tutorial review,'' \emph{IEEE Open Journal of the Communications Society}, vol.~4, pp. 1999--2049, 2023.

\bibitem{10382465}
Y.~Chen, H.~Hua, J.~Xu, and D.~W.~K. Ng, ``Isac meets swipt: Multi-functional wireless systems integrating sensing, communication, and powering,'' \emph{IEEE Transactions on Wireless Communications}, vol.~23, no.~8, pp. 8264--8280, 2024.

\bibitem{10812799}
Y.~Xu, D.~Xu, and S.~Song, ``Sensing-assisted robust swipt for mobile energy harvesting receivers in networked isac systems,'' \emph{IEEE Transactions on Wireless Communications}, pp. 1--1, 2024.

\bibitem{10382692}
H.~Lin, M.~A. Kishk, and M.-S. Alouini, ``Performance evaluation of rf-powered iot in rural areas: The wireless power digital divide,'' \emph{IEEE Transactions on Green Communications and Networking}, vol.~8, no.~2, pp. 716--729, 2024.

\bibitem{10711847}
H.~K. Armeniakos, P.~S. Bithas, K.~Maliatsos, and A.~G. Kanatas, ``Joint energy and sinr coverage probability in uav corridor-assisted rf-powered iot networks,'' \emph{IEEE Communications Letters}, vol.~28, no.~12, pp. 2904--2908, 2024.

\bibitem{10578306}
N.~G. Evgenidis, N.~A. Mitsiou, V.~I. Koutsioumpa, S.~A. Tegos, P.~D. Diamantoulakis, and G.~K. Karagiannidis, ``Multiple access in the era of distributed computing and edge intelligence,'' \emph{Proceedings of the IEEE}, vol. 112, no.~9, pp. 1497--1526, 2024.

\bibitem{liu2022learning}
C.~Liu, W.~Yuan, S.~Li, X.~Liu, H.~Li, D.~W.~K. Ng, and Y.~Li, ``Learning-based predictive beamforming for integrated sensing and communication in vehicular networks,'' \emph{IEEE Journal on Selected Areas in Communications}, vol.~40, no.~8, pp. 2317--2334, 2022.

\bibitem{10113889}
X.~Liu, H.~Zhang, K.~Long, A.~Nallanathan, and V.~C.~M. Leung, ``Distributed unsupervised learning for interference management in integrated sensing and communication systems,'' \emph{IEEE Transactions on Wireless Communications}, vol.~22, no.~12, pp. 9301--9312, 2023.

\bibitem{10229204}
Y.~Liu, I.~Al-Nahhal, O.~A. Dobre, F.~Wang, and H.~Shin, ``Extreme learning machine-based channel estimation in irs-assisted multi-user isac system,'' \emph{IEEE Transactions on Communications}, vol.~71, no.~12, pp. 6993--7007, 2023.

\bibitem{mao2024uav}
W.~Mao, Y.~Lu, G.~Pan, and B.~Ai, ``Uav-assisted communications in sagin-isac: Mobile user tracking and robust beamforming,'' \emph{IEEE Journal on Selected Areas in Communications}, 2024.

\bibitem{favarelli2025sensor}
E.~Favarelli, E.~Matricardi, L.~Pucci, W.~Xu, E.~Paolini, and A.~Giorgetti, ``Sensor fusion and resource management in mimo-ofdm joint sensing and communication,'' \emph{IEEE Transactions on Vehicular Technology}, 2025.

\bibitem{bithas2024joint}
P.~S. Bithas, G.~P. Efthymoglou, A.~G. Kanatas, and K.~Maliatsos, ``Joint sensing and communications in unmanned-aerial-vehicle-assisted systems,'' \emph{Drones}, vol.~8, no.~11, p. 656, 2024.

\bibitem{pradhan2023stochastic}
A.~Pradhan, H.~S. Dhillon, F.~Tufvesson, and A.~F. Molisch, ``Stochastic geometry analysis of a new gscm with dual visibility regions,'' in \emph{2023 IEEE 34th Annual International Symposium on Personal, Indoor and Mobile Radio Communications (PIMRC)}.\hskip 1em plus 0.5em minus 0.4em\relax IEEE, 2023, pp. 1--6.

\bibitem{10716723}
D.~Tyrovolas, D.~Bozanis, S.~A. Tegos, V.~K. Papanikolaou, P.~D. Diamantoulakis, C.~K. Liaskos, R.~Schober, and G.~K. Karagiannidis, ``Empowering programmable wireless environments with optical anchor-based positioning,'' \emph{IEEE Network}, vol.~39, no.~1, pp. 14--20, 2025.

\bibitem{bozanis2024}
\BIBentryALTinterwordspacing
D.~Bozanis, D.~Tyrovolas, V.~K. Papanikolaou, S.~A. Tegos, P.~D. Diamantoulakis, C.~K. Liaskos, R.~Schober, and G.~K. Karagiannidis, ``Location-driven programmable wireless environments through light-emitting ris ({LeRIS}),'' 2024. [Online]. Available: \url{https://arxiv.org/abs/2412.04989}
\BIBentrySTDinterwordspacing

\end{thebibliography}

\end{document}